\documentclass[journal=jctcce,manuscript=article]{achemso}

\usepackage{amsmath}
\usepackage{amssymb}
\usepackage{graphicx}
\usepackage{xcolor}
\usepackage{booktabs}
\usepackage{multirow}
\usepackage[version=4]{mhchem}
\usepackage[hidelinks]{hyperref}
\newcommand{\RMSD}{\text{RMSD}}
\newcommand{\MSD}{\text{MSD}}
\newcommand{\neff}{n_\text{eff}}
\newcommand{\sigop}{\sigma_\text{op}}
\newcommand{\taup}{\tau_\text{op}}
\newcommand{\bvec}[1]{\boldsymbol{#1}}

\author{Saumyak Mukherjee}
\affiliation{Theoretical Biophysics, Max Planck Institute of Biophysics, Frankfurt am Main, Germany}
\author{Gerhard Hummer}
\affiliation{Theoretical Biophysics, Max Planck Institute of Biophysics, Frankfurt am Main, Germany}
\alsoaffiliation{Institute of Biophysics, Goethe University Frankfurt, Frankfurt am Main, Germany}
\email{gerhard.hummer@biophys.mpg.de}

\title{Bayesian Learning of Distance Metrics Beyond RMSD for Biomolecule Alignment, Clustering, and Domain Identification}

\begin{document}

\begin{abstract}
Root-mean-square deviation (RMSD) is the standard metric of structural comparison in molecular dynamics (MD) simulations. In its conventional form, RMSD assigns equal weight to all atoms regardless of mobility. Hence, flexible loops and disordered regions can dominate a global RMSD, while the rigid functional core contributes negligibly to the overall metric. To address this issue, we introduce the Bayes-optimal RMSD (\emph{BRMSD}), which optimizes per-atom weights jointly with structural averages by maximizing a Bayesian posterior. In a trade-off between low RMSD and weight uniformity, a position-fluctuation parameter $\sigma$ controls the transition from classical RMSD ($\sigma \to \infty$) to a progressive focus on a rigid core ($\sigma \to 0$). The BRMSD framework supports analysis modules for structural alignment, focused alignment onto a user-specified domain, trajectory smoothing, soft $K$-means conformational clustering, and rigid-domain identification. These modules are implemented in the open-source Python package \href{https://github.com/bio-phys/BRMSD}{\texttt{BRMSD}} and benchmarked on two MD systems, the endoplasmic reticulum translocon-associated protein SND3 and the phosphotransferase adenylate kinase.
\end{abstract}

\section{Introduction}

Root-mean-square deviation (RMSD) is a fundamental metric, universally used for quantifying structural similarity in MD simulations.\cite{hayward2023, kabsch1976, kabsch1978, theobald2005, maiorov1994} Its utility comes from interpretability. It has units of length and a well-defined physical meaning as a measure of distance in conformation space. RMSD underlies convergence assessment, free-energy landscapes, conformational-state identification, and domain-motion analysis, all of which rely on the spatial deviation between conformations. The standard definition of RMSD assigns equal weight to a pre-selected set of atoms in a system, say the C$_\alpha$ atoms of a protein backbone. However, this a priori choice may conflate structurally informative and uninformative atoms. For example, in proteins with flexible loops, disordered termini, or mobile linkers, a handful of highly mobile residues can dominate the RMSD while focusing marginally on the conformational state of the functional core. Workarounds involve manual atom selection or iterative exclusion of outliers. \cite{grossfield2019} However, it is difficult to generalize these approaches since they are user-dependent and require prior structural knowledge. This compromises reproducibility as MD simulation campaigns become increasingly automated to cover a wide range of molecular systems and states.

Rather than relying on ad hoc selection rules, a cleaner approach is to treat atomic weights as active variables within the structural alignment optimization. Early protocols implemented static assignments, such as McLachlan's mass-weighted superposition~\cite{mclachlan1979}, which remain blind to the fluctuations of a given molecular dynamics trajectory. Later ensemble-based methodologies introduced by Damm and Carlson~\cite{damm2006} (via Gaussian variance scaling) and Theobald and Wuttke~\cite{theobald2008} (via maximum-likelihood inference) succeeded in capturing trajectory-specific dynamics. However, these methods treat weight calculation and coordinate superposition as decoupled, sequential operations, leaving open the question of mathematical self-consistency between the weights and the resultant average structure.

Here, we bridge this gap by consolidating weight optimization and structural alignment into a single free-energy functional. In a Bayesian formulation, we derive the free energy as the negative logarithm of the Bayes posterior for the atom weights and reference structure. Minimizing the free energy then corresponds to a multi-objective optimization process that balances the competing demands to align the structures onto a quasi-rigid core and to include as many atoms into this core as possible. We refer to the RMSD calculated with optimized weights as ``Bayes-optimal RMSD'' (BRMSD). Equivalently, this trade-off between low mean-squared deviation (MSD) and near-uniform weights is achieved by combining the usual metric of MSD with a relative entropy regularizer that penalizes non-uniform atom weight distributions.\cite{jaynes1957,rozycki_saxs_2011} Analogous approaches have been applied in reweighting and modeling of MD ensembles\cite{rozycki_saxs_2011, hummer2015, pitera2012, bottaro2020, gilardoni2026, bonomi2016, frohling2023}.

We minimize a free-energy functional $G$ to achieve low MSD while simultaneously penalizing deviations from the normalized prior weights $W_\alpha\ge 0$ with $\sum_\alpha W_\alpha=1$. Here, we assume uniform prior weights, $W_\alpha = 1/N$, where $N$ is the total number of atoms. As penalty, we add the Kullback--Leibler (KL) divergence term,\cite{Kullback1951} $S_\mathrm{KL}(\bvec{w} \| \bvec{W}) = \sum_\alpha w_\alpha \ln(w_\alpha / W_\alpha)$,  to the total MSD from a reference structure. Simultaneous minimization over structure, rotations, and weights yields a fixed-point iteration that converges to the entropy-optimal weights.

The weight functional suppresses atoms in proportion to their mean-square fluctuations about the average structure, concentrating weight on rigid atoms and down-weighting flexible ones without the requirement of manual selection. The same functional form, adapted to incorporate window weights, domain bias, cluster assignments, or inter-domain overlap penalties, can be adapted to multiple distinct tasks, including structural alignment, focused alignment, trajectory smoothing, soft $K$-means clustering with cluster-specific atom weights, and rigid domain detection. All five modules share a single position-fluctuation parameter $\sigma$ in units of {\AA}. Large values of $\sigma$ keep the KL divergence dominant, holding weights near the uniform prior and recovering classical RMSD. Smaller values of $\sigma$ allow the concentration of weight onto the most rigid atoms.

We benchmarked BRMSD on two systems chosen to represent complementary structural regimes. Alignment and smoothing analysis were performed on a 300~ns all-atom MD trajectory of SND3\cite{yang2025, Zenodo2025}, an eukaryotic endoplasmic reticulum translocon-associated membrane protein, featuring a distinct transmembrane groove and flexible cytosolic claw domains. Clustering and rigid domain identification analyses were performed on the adenylate kinase (ADK) protein, which contains three distinct domains and has been shown to transition between open and closed states\cite{beckstein2009}. For this analysis we use the directed transition trajectories distributed through the MDAnalysisData repository. \cite{mda2011, mda2016, mdanalysisdata}

\section{Theory}

\subsection{Weighted Mean-Square Deviation}

Let $x_i=(r_{i1},r_{i2},\ldots,r_{iN})$ denote the $3N$-dimensional vector containing the three-dimensional Cartesian coordinates $r_{i\alpha}$ of the $N$ atoms in molecular structure $i$. Given a normalized atom weight vector $\bvec{w} = \{w_\alpha\}\;(\alpha = 1, \ldots, N)$ with $\sum_\alpha w_\alpha = 1$ and $w_\alpha \ge 0$, the weighted mean square deviation (\MSD) between two structures after optimal rigid-body superposition is given by 
\begin{equation}
  \MSD(x_1, x_2; \bvec{w}) = \min_R \sum_\alpha w_\alpha
    \lvert r_{1\alpha} - R\, r_{2\alpha} \rvert^2,
  \label{eq:msd}
\end{equation}
Here $R \in \mbox{SE}(3)$ denotes a rigid-body transformation. As a member of the special Euclidean group, $R$ consists of a translation (of the entire structure) and a rotation (about its center, as defined by the weighted average of atom positions with weights $\alpha$). $|\cdot|$ represents the 3D Euclidean distance. The optimal $R$ for a given weight vector $\bvec{w}$ is found by the Kabsch algorithm.\cite{kabsch1976, kabsch1978} The root-mean-square deviation is $\RMSD = \MSD^{1/2}$. Classical RMSD corresponds to the uniform prior $w_\alpha = W_\alpha \equiv 1/N$ for all $\alpha$.

\subsection{Bayesian Formulation of RMSD Structural Alignment}
In a Bayesian formulation of RMSD structural alignment, we assume a priori that configurations $x$ in $3N$-dimensional configuration space are distributed as
\begin{equation}
  \label{eq:pofx}
  p_0(x|\boldsymbol{w},s,\sigma)\propto \exp\left(-\frac{\mbox{MSD}(x,s;\boldsymbol{w})}{2\sigma^2} \right)
\end{equation}
where $s$ is the reference structure and $\sigma$ is the expectation of the root-mean-squared fluctuation per atom. As a Bayes prior for the weight vector, we assume that the optimal atom weight vector $\boldsymbol{w}$ entering the $\mbox{MSD}$ is distributed as
\begin{equation}
  \label{eq:pofw}
  q_0(\boldsymbol{w}|\boldsymbol{W},\theta)\propto \exp(-\theta\, S_{\mathrm{KL}}(\boldsymbol{w}||\boldsymbol{W}))
\end{equation}
where $\boldsymbol{W}$ is a vector of reference weights and $\theta$ is a confidence parameter. Analogous to Bayesian ensemble refinement \cite{hummer2015}, we then use the Bayes relation to define a Bayes posterior for $s$ and $\boldsymbol{w}$ given configurations $x_1,x_2,\ldots$,
\begin{align}
  \label{eq:posterior}
  &\mbox{posterior}(s,\boldsymbol{w}|x_1,x_2,\ldots;\boldsymbol{W},\theta,\sigma)\nonumber\\
  &\propto\prod_i p_0(x_i|\boldsymbol{w},s,\sigma)q_0(\boldsymbol{w}|\boldsymbol{W},\theta)=\exp(-g)
\end{align}
with an effective free energy
\begin{align}
  \label{eq:gsmall}
  &g=\sum_i\frac{\mbox{MSD}(x_i,s;\boldsymbol{w})}{2\sigma^2}
  +\theta\, S_{\mathrm{KL}}(\boldsymbol{w}||\boldsymbol{W})
\end{align}
The optimal reference and weight vector are then defined by maximizing the Bayes posterior, or equivalently by minimizing the effective free energy,
\begin{equation}
  \label{eq:maxposterior}
  (s,\boldsymbol{w})= \mbox{argmax}_{s,\boldsymbol{w}}\,\mbox{posterior}(s,\boldsymbol{w}) = \mbox{argmin}_{s,\boldsymbol{w}}\,g
\end{equation}
Maximizing the posterior is unchanged if we multiply $g$ by the constant $2\sigma^2$. Setting the prior confidence to $\theta = M/2$ fixes the entropy penalty at a per-frame weight $\sigma^2$, so its total weight becomes $M\sigma^2$. With this choice the confidence parameter is no longer free, and we minimize the effective free energy
\begin{equation}
  \label{eq:G_align}
  G=\sum_i\mbox{MSD}(x_i,s;\boldsymbol{w})
  + M\sigma^2 \sum_\alpha w_\alpha\ln\frac{w_\alpha}{W_\alpha}
\end{equation}
with respect to $s$ and $\boldsymbol{w}$. In the following, we  use $\sigma$ or
\begin{equation}
\label{eq:theta}
\theta=M\sigma^2
\end{equation}
as fully equivalent parameters to control the trade-off between minimizing the $\mbox{MSD}$ and the KL divergence. From here on $\theta$ denotes $M\sigma^2$, not the prior confidence of eq~\ref{eq:pofw}. With this Bayesian perspective in mind, we refer to the RMSD alignment procedure using the optimal reference structure $s$ and optimal weights $\boldsymbol{w}$ as ``Bayesian inference of RMSD'' (BRMSD).

We note the close analogy to an expression derived originally in the ``ensemble refinement of SAXS'' (EROS) method \cite{rozycki_saxs_2011} and then worked out in detail in ref \citenum{hummer2015}. Here, the MSD term replaces the $\chi^2$ error of ensemble refinement. As in ensemble refinement, an expression identical to eq \ref{eq:G_align} is obtained by combining the usual metric of MSD with a relative entropy regularizer to penalize non-uniform atom weight distributions.\cite{jaynes1957, rozycki_saxs_2011, hummer2015}

\subsection{Structural Alignment}

To find the weighted average structure $s$ of an ensemble of $M$ structures $\{x_i\}$, we minimize $G$ over the average structure $s$, the rigid-body superpositions $\{R_i\}$ of $x_i$ and $s$, the reference structure $s$, and the atom weights $\bvec{w}$, subject to the normalization condition of the weights. The first term in $G$, as given in eq \ref{eq:G_align}, is the aggregate mean-square deviation from the average structure. The second term represents the relative entropy of $\bvec{w}$ with respect to the uniform prior $W_\alpha = 1/N$, scaled by $\theta = M\sigma^2$. 

The minimization is achieved by setting derivatives of $G$ with respect to $s_\alpha$ and $w_\alpha$ to zero. This yields the self-consistency equations
\begin{align}
  \label{eq:w_update}
  w_\alpha^{(n+1)} &\propto W_\alpha \exp \left(-\frac{1}{M\sigma^2}\displaystyle\sum_i \lvert s_\alpha^{(n)} - R_i^{(n)} r_{i\alpha} \rvert^2\right),\\
    s_\alpha^{(n+1)} &= \frac{1}{M} \sum_{i=1}^M R_i^{(n+1)} r_{i\alpha},
  \label{eq:s_update}
\end{align}
where $R_i^{(n)}$ is the optimal rigid-body translation and rotation of $x_i$ for weights $\bvec{w}^{(n)}$ with $s^{n}$ as reference, $R_i^{(n)}r_{i\alpha}$ its application to atom $\alpha$ of $x_i$,
and the weights $w_\alpha^{(n)}$ are normalized after each iteration step $n$. Starting from $s^{(1)} = x_1$ and $w_\alpha^{(1)} = W_\alpha$, eqs~\ref{eq:s_update} and \ref{eq:w_update} are iterated towards convergence.

$G$ has a lower bound of zero since the MSD term is a sum of squared norms and $S_\mathrm{KL}(\bvec{w}\|\bvec{W})\ge 0$ by Gibbs' inequality\cite{jaynes1957}. Each alternating minimization step decreases $G$ monotonically. The weight update minimizes $G$ exactly at fixed $s$, and the structure update minimizes the MSD term at fixed $\bvec{w}$. Such a decrease of a function bounded from below guarantees convergence to a stationary point. Since $G$ is non-convex, this point may be a local, rather than a global minimum. Therefore, although for alignment and smoothing, a single initialization from $s^{(1)} = x_1$ and $w_\alpha^{(1)} = W_\alpha$ may suffice, for clustering and domain identification, we run $n_r$ random restarts and retain the solution with the lowest $G$.

The sum in the exponent of eq~\ref{eq:w_update} is the per-atom mean-square fluctuation (MSF) about the current average. Accordingly, atoms with large fluctuations receive exponentially suppressed weights, while structurally rigid atoms are up-weighted. Atoms whose summed squared deviation $\sum_i |s_\alpha - R_i r_{i\alpha}|^2$ exceeds $\theta = M\sigma^2$ are exponentially suppressed, whereas those below retain near-uniform weight. Equivalently, an atom is suppressed once its mean-square fluctuation averaged over the $M$ frames exceeds $\sigma^2$, so the weights are set by the per-atom mean-square fluctuation rather than by the number of frames $M$ used to estimate it, provided the sampling is converged. At $\sigma \to \infty$ the entropy term dominates and weights stay uniform, recovering classical RMSD. The optimal $\sigma$ depends on the module and on the fluctuation amplitude of the system.

The entropy-based effective atom count is the exponential of the Shannon entropy of the weight distribution (eq~\ref{eq:neff}),
\begin{equation}
  \neff = \exp\!\left(-\sum_\alpha w_\alpha \ln w_\alpha\right)
  \label{eq:neff}
\end{equation}
where, $\neff$ equals $N$ for uniform weights and decreases monotonically as weight concentrates. 
\subsection{Focused Alignment}

The BRMSD alignment concentrates weight on the most globally rigid regions of the protein, such as the TMD groove in SND3. However, the most functionally relevant regions may not always coincide with the rigid structural core. For instance, a flexible domain critical for substrate binding will warrant non-negligible weight alongside the globally rigid scaffold.  Hence, depending on the problem at hand, one may wish to focus the alignment on a given domain of interest $D$ while still including the remainder of the structure. We augment the free-energy functional of eq~\ref{eq:G_align} with a concentration penalty on atoms in $D$ (eq~\ref{eq:G_focus}).
\begin{equation}
  G = \sum_i \MSD(x_i, s; \bvec{w})
    + \theta \sum_\alpha w_\alpha \ln \frac{w_\alpha}{W_\alpha}
    + \mu \sum_{\alpha \in D} w_\alpha \ln(n_D\,w_\alpha),
  \label{eq:G_focus}
\end{equation}
where $n_D$ is the number of atoms in $D$ and $\mu > 0$ controls the strength of the concentration bias toward $D$. Taking the derivative with respect to $w_\alpha$ yields two separate update rules.
\begin{align}
  w_\alpha^{(n+1)} &\propto W_\alpha \exp\!\left(
    -\frac{\sum_i \lvert s_\alpha^{(n)} - R_i^{(n)} r_{i\alpha} \rvert^2}{\theta}
    \right), \quad \alpha \notin D,
  \label{eq:w_focus_out} \\
  w_\alpha^{(n+1)} &\propto
    \frac{W_\alpha^{\theta/(\theta+\mu)}}{n_D^{\mu/(\theta+\mu)}}
    \exp\!\left(
    -\frac{\sum_i \lvert s_\alpha^{(n)} - R_i^{(n)} r_{i\alpha} \rvert^2}{\theta+\mu}
    \right), \quad \alpha \in D.
  \label{eq:w_focus_in}
\end{align}
Atoms outside $D$ follow the standard alignment update (with equivalent eq~\ref{eq:w_focus_out} and eq~\ref{eq:w_update}), whereas atoms inside $D$ experience a reduced effective concentration parameter $\theta + \mu$ and an additional prior bias $W_\alpha^{\theta/(\theta+\mu)}$ that amplifies their weight relative to atoms outside $D$. In the limit $\mu \gg \theta$, the atoms in $D$ receive weight $w_\alpha \propto n_D^{-1}$ (uniform within $D$) regardless of their MSF. As before, all weights are renormalized after each update, and the structure update follows eq~\ref{eq:s_update}.

The dimensionless ratio $\mu/\theta$ controls the degree of domain focusing. A practical criterion for selecting the operating point $(\mu/\theta)_\text{op}$ is that $\neff$ should remain above $\lvert D \rvert$. If $\neff < \lvert D \rvert$, the alignment is effectively performed on fewer atoms than the domain contains, which can cause numerical instability.

\subsection{Trajectory Smoothing}

Structural alignment averages uniformly over all $M$ frames. High-frequency positional noise from thermal fluctuations is thereby retained in each individual frame, even after alignment. To suppress this noise while preserving slow conformational dynamics, we replace the global average with a local one centered on each frame $j$. We introduce frame-dependent weights $p(i|j) \ge 0$, $\sum_i p(i|j) = 1$, that localize structure $i$ relative to frame $j$. A triangular window of half-width $W$ sets $p(i|j) \propto \max(0, W - |i-j|)$ or a uniform window can be used. The smoothed-frame functional replaces eq~\ref{eq:G_align} by
\begin{equation}
  G_j = \sum_i p(i|j)\,\MSD(x_i, s_j; \bvec{w}_j)
    + \theta \sum_\alpha w_{j\alpha} \ln \frac{w_{j\alpha}}{W_\alpha},
  \label{eq:G_smooth}
\end{equation}
minimized independently for each frame $j$. The fixed-point equations are identical in form to eqs~\ref{eq:s_update}--\ref{eq:w_update} with the uniform sum replaced by the windowed sum $\sum_i p(i|j)$. Because the window weights are normalized ($\sum_i p(i|j) = 1$), this windowed sum is a per-frame average, so the entropy penalty here uses $\theta = \sigma^2$ (the effective frame count is $M = 1$). The converged $s_j$ is the smoothed coordinate of frame $j$, and the collection $\{s_j\}$ constitutes the smoothed trajectory.

\subsection{Conformational Clustering}

Structural alignment identifies a single average structure and a single weight vector for the full ensemble. When the ensemble spans multiple conformational states, a single average is insufficient. We extend the alignment framework to $K$ cluster centers by replacing the uniform frame average with a soft assignment. Each frame $i$ is assigned to cluster $a$ with responsibility\cite{mackay2003} $q(a|i) \ge 0$, $\sum_a q(a|i) = 1$. Each cluster carries its own center $s_a$ and, crucially, its own atom weight vector $\bvec{w}_a$, so the alignment metric adapts to the internal rigidity pattern of each conformational state rather than imposing a single global metric. The joint free-energy functional is given by eq~\ref{eq:G_cluster}.
\begin{equation}
  G = \sum_a \sum_i q(a|i) \sum_\alpha w_{a\alpha}
        \lvert s_{a\alpha} - R_{ia} r_{i\alpha} \rvert^2
    + \theta \sum_a \sum_\alpha w_{a\alpha} \ln \frac{w_{a\alpha}}{W_\alpha}
    + \tau   \sum_i \sum_a q(a|i) \ln \bigl(M\,q(a|i)\bigr),
  \label{eq:G_cluster}
\end{equation}
Here, the third term penalizes deterministic assignments (large $\tau$ encourages soft, diffuse responsibilities, small $\tau$ recovers hard $K$-means). Setting $\partial G/\partial q(a|i) = 0$ subject to $\sum_a q(a|i) = 1$ via a Lagrange multiplier yields the softmax responsibility update (eq~\ref{eq:q_update}), where $s_a$ is the center of cluster $a$, $\theta$ controls atom weight concentration (eq~\ref{eq:theta}), and $\tau > 0$ (units \AA$^2$) controls cluster assignment softness. The fixed-point iteration is
\begin{align}
  q^{(n+1)}(a|i) &\propto \exp\!\left(
    -\frac{\sum_\alpha w_{a\alpha}^{(n)}
      \lvert s_{a\alpha}^{(n)} - R_{ia}^{(n)} r_{i\alpha}\rvert^2}{\tau}\right),
  \label{eq:q_update} \\
  w_{a\alpha}^{(n+1)} &\propto W_\alpha\exp\!\left(
    -\frac{\sum_i q^{(n+1)}(a|i)
      \lvert s_{a\alpha}^{(n)} - R_{ia}^{(n)} r_{i\alpha}\rvert^2}{\theta}\right),
  \label{eq:wa_update} \\
  s_{a\alpha}^{(n+1)} &=
    \frac{\sum_i q^{(n+1)}(a|i)\,R_{ia}^{(n+1)} r_{i\alpha}}
         {\sum_i q^{(n+1)}(a|i)}.
  \label{eq:sa_update}
\end{align}
The cluster center $s_{a\alpha}$ is the responsibility-weighted average structure and represents the fixed point that minimizes the weighted MSD across all frames. Each cluster has its own weight vector $\bvec{w}_a$, so the alignment metric adapts to each cluster's internal rigidity pattern.

\subsection{Rigid Domain Identification}

Conformational clustering partitions frames into groups (clusters), whereas rigid domain identification partitions atoms into groups (domains). Both share the BRMSD free-energy structure, but the variable of interest shifts from frame assignments $q(a|i)$ to atom weight vectors $\bvec{w}_k$.

The identification of rigid domains is done by sequential peeling of atoms. A series of independent optimizations with $K=1$ domain is performed on shrinking atom pools. At each step, the BRMSD functional (eq~\ref{eq:G_align}) is minimized on the current pool to obtain a single weight vector $\bvec{w}$ concentrated on the most rigid sub-structure within that pool. Atoms with $w_\alpha > t \cdot \max_\beta w_\beta$ are assigned to the current domain and removed from the pool. The threshold $t$ controls pool composition for subsequent steps. Small values over-claim atoms, depleting later pools, and large values underclaim, leaving rigid residues in the pool that can dominate subsequent steps. The procedure repeats on the remaining atoms until no atoms satisfy the threshold or $K$ domains have been recovered. The peeling order is data-determined, such that the most globally rigid sub-structure is identified first. Hence, domains emerge in order of decreasing rigidity. At each step, $\theta = M\sigma^2$ is held fixed and independent of the pool size, preserving the physical meaning of $\sigma$ as the per-atom root-mean-squared fluctuation (RMSF) threshold across all steps. Domain detection quality is assessed by the Jaccard index $J(A,B) = |A \cap B|/|A \cup B|$ between each recovered domain and the reference assignment.

An alternative formulation seeks $K$ non-overlapping weight vectors simultaneously, enforcing mutual exclusivity through an overlap penalty on the dot products $\bvec{w}_j \cdot \bvec{w}_k$.
\begin{equation}
  G = \sum_k \sum_i \MSD(x_i, s_k; \bvec{w}_k)
    + \theta \sum_k \sum_\alpha w_{k\alpha} \ln \frac{w_{k\alpha}}{W_\alpha}
    + \frac{\mu}{2} \sum_{j < k} \left(\bvec{w}_j \cdot \bvec{w}_k\right)^2.
  \label{eq:G_domain}
\end{equation}
The weight update becomes
\begin{equation}
  w_{k\alpha}^{(n+1)} \propto W_\alpha \exp\!\left(
    -\frac{\sum_i \lvert s_{k\alpha}^{(n)} - R_{ki}^{(n)} r_{i\alpha}\rvert^2}{\theta}
    - \mu \sum_{j \ne k} \bigl(\bvec{w}_j^{(n)} \cdot \bvec{w}_k^{(n)}\bigr)
      w_{j\alpha}^{(n)} \right).
  \label{eq:wk_update}
\end{equation}
This update is obtained by differentiating $G$ with respect to $w_{k\alpha}$, adding a Lagrange multiplier for the simplex constraint $\sum_\alpha w_{k\alpha} = 1$, and setting the derivative to zero. The overlap gradient $\mu\sum_{j\ne k} (\bvec{w}_j\cdot\bvec{w}_k)w_{j\alpha}$ enters additively in the exponent, coupling domain $k$'s weights to all other domains. Since the coupling is simultaneous, the weight updates for different $k$ are not independent. In practice, simultaneous optimization can fail when the overlap-penalty gradient is too weak to prevent weight vectors from collapsing onto the globally most rigid region (see Results). Sequential peeling resolves this by construction.

Implementation details, including timing benchmarks for all modules (Table~\ref{tab:timing}), are provided in the Supporting Information.

\subsection{Extension to Mahalanobis-Like RMSD Metric}
For completeness, we also sketch an extension to a Mahalonobis-like RMSD that accounts for anisotropic fluctuations in atom positions. Let us assume that we have an initial alignment, which gives us the aligned structures $x_i$. In a form of PCA,\cite{Garcia:PRL:1992} we then construct the average $s=\sum_{i=1}^Mx_i/M$ of the $M$ structures and the covariance matrix $C=\sum_{i=1}^M\delta x_i\delta x_i^T/M$ with $\delta x_i=x_i-s$ and $T$ the transpose. A spectral decomposition of $C$ (or a singular value decomposition of the $x_i$) then gives us $n$ eigenvalues $\lambda_i>0$ with orthonormal eigenvectors $v_i$, i.e., $v_i^Tv_j=\delta_{ij}$ with $\delta_{ij}$ the Kronecker delta. At least six of the eigenvalues will be zero, corresponding to rigid body translations and rotations, $n\le 3N-6$. We define the free energy by giving weights $w_j$ to the PCA modes $v_j$ as
\begin{equation}
  \label{eq:GPCA}
  G=\sum_{i=1}^M\sum_{j=1}^nw_j[v_j^T(\boldsymbol{R}x_i-s)]^2+\theta\sum_{j=1}^nw_j\ln\frac{w_j}{W_j}
\end{equation}
where $\boldsymbol{R}=I_N\otimes R$ applies the translation and rotation  atom-wise to the configurations. Minimization with respect to $s$ gives the average position, $s=\sum_{i=1}^M\boldsymbol{R}x_i/M$, as before. Setting the derivative with respect to the normalized weights $w_j$ to zero gives us an update rule for the mode weights,
\begin{equation}
  \label{eq:wjPCA}
  w_j\propto W_j\,e^{-\sum_{i=1}^M[v_j^T(\boldsymbol{R}x_i-s)]^2/\theta}
\end{equation}
However, we have not currently implemented this procedure because the rigid-body alignment code to optimize $R$ will have to be adapted to the more general metric. After iteration to self-consistency, one would then interpolate between an alignment according to the Mahalanobis distance ($\theta=0$) and the usual Euclidean distance ($\theta\to\infty$).

\section{Results and Discussion}

\subsection{Systems}

We benchmarked BRMSD on two systems that represent complementary challenges. For alignment and smoothing, we used SND3, which is a 191-residue eukaryotic translocon-associated protein recently resolved by cryo-EM at 3.1~\AA{} resolution (PDB: 9I78).\cite{yang2025} Three transmembrane helices (TMD-2, residues 28--60; TMD-3, residues 86--115; TMD-4, residues 119--134) form a buried TM groove that constitutes the rigid functional core. An additional, shorter transmembrane helix (TMD-1, residues 5--21) is embedded in the membrane and is less rigid than the other three TMDs (it serves as the target domain for the focused alignment benchmark). Two cytosolic claw domains (N-claw, residues 61--85; C-claw, residues 136--173) and a C-terminal tail (residues 174--191) are substantially more mobile. We analyzed a 300~ns all-atom MD trajectory of the isolated protein chain, sampled at 100~ps intervals. The SND3 trajectory samples equilibrium fluctuations around a single mean structure without discrete conformational states.

The conformational clustering and domain identification were benchmarked on adenylate kinase (ADK), a phosphotransferase that undergoes a large-scale open-to-closed conformational transition upon substrate binding~\cite{beckstein2009}. Crystallography and mutational studies have established three structurally distinct regions in ADK to single-residue precision, namely the CORE (residues 1--29, 60--121, and 160--214), NMPbind (residues 30--59), and LID (residues 122--159) domains.\cite{beckstein2009} The two endpoint crystal structures, open (PDB ID: 4AKE) and closed (PDB ID: 1AKE), are characterized by NMPbin--LID distances of 32.1 and 19.6~\AA{}, and LID--CORE distances of 40.0 and 30.0~\AA{}, respectively. To evaluate clustering performance, we utilized 200 independent Dynamic Importance Sampling (DIMS) trajectories~\cite{seyler2014} (with a total of 19,691 frames) that span this open-to-closed transition. This ensemble represents a non-equilibrium transition rather than a continuous equilibrium trajectory. For domain detection validation, we employed 28,282 Framework Rigidity Optimized Dynamics Algorithm (FRODA) frames~\cite{wells2005}. FRODA propagates rigid-body constraints on covalent geometry, so intra-domain RMSF is substantially lower than inter-domain RMSF by construction.

\subsection{Alignment}

\begin{figure}[ht!]
  \centering
  \includegraphics[width=\linewidth]{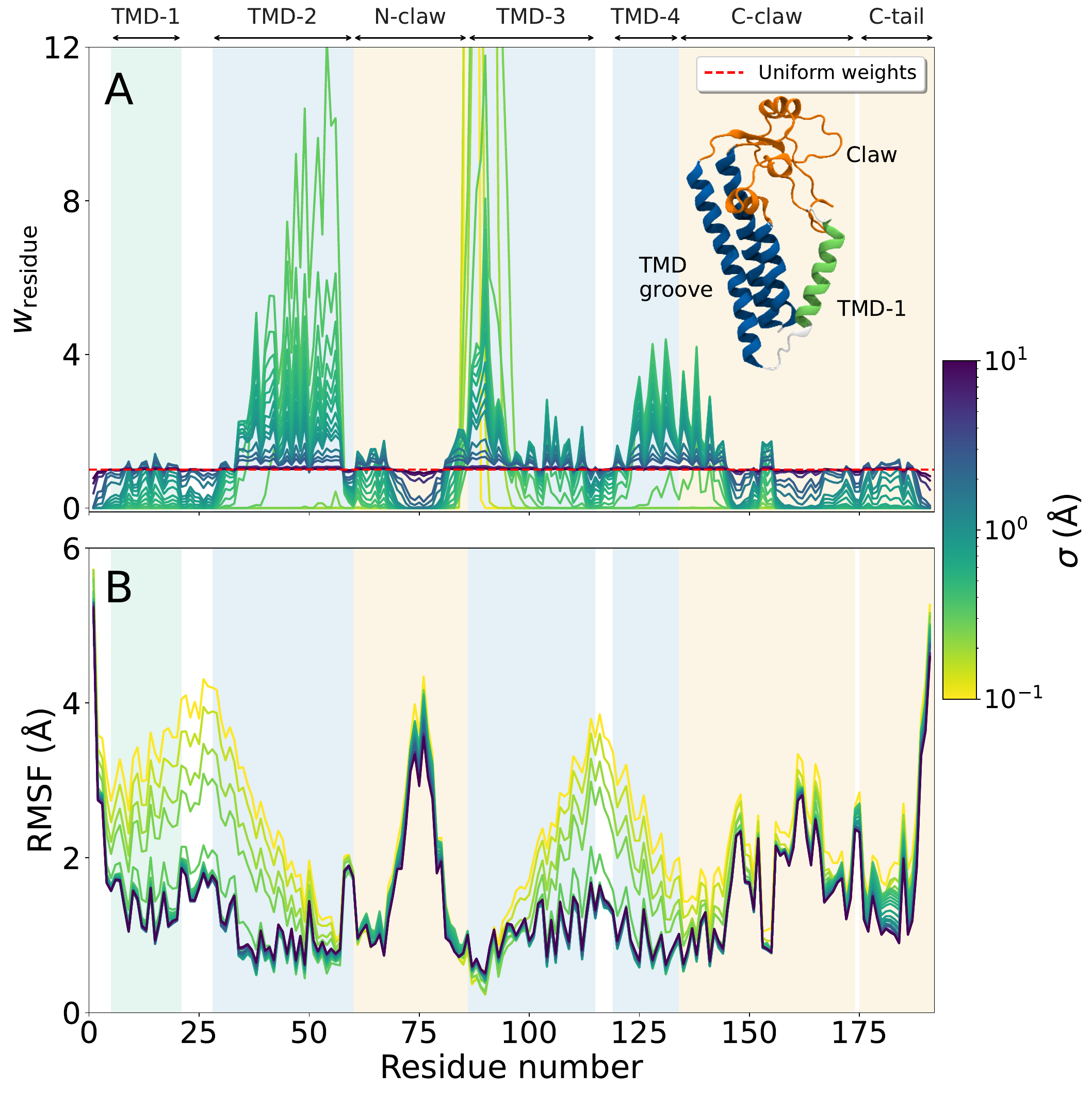}
  \caption{The $\sigma$ sweep for SND3 structural alignment is shown in two panels. (A) Per-residue weight $w_{\rm residue}$ (mean atom weight of a residue, scaled so a uniform weight equals one) as a function of residue number for $\sigma = 0.1$--10.0~\AA{}, colored by $\sigma$ (colorbar). Domain shading indicates TMD-1 (green), TMD-2/TMD-3/TMD-4 (blue, TM groove), and N-claw/C-claw/C-tail (orange). The SND3 structure (PDB: 9I78), colored according to the domains, is shown as an inset. As $\sigma$ decreases, weight concentrates progressively in the TM groove. (B) Per-residue RMSF for the same $\sigma$ values. The profile is independent of $\sigma$ (other than at very low $\sigma$, due to over-regularization), confirming that the weight redistribution does not alter the underlying fluctuations.}
  \label{fig:snd3_weight}
\end{figure}

We scanned $\sigma$ from 0.1 to 10~\AA{} on the SND3 trajectory (Figure~\ref{fig:snd3_weight}). As $\sigma$ decreases, the per-residue weights $w_{\rm residue}$ grow monotonically in the TM groove (TMD-2, TMD-3, TMD-4) while collapsing toward zero in the more mobile regions (Figure~\ref{fig:snd3_weight}A). The per-residue RMSF profile is mostly insensitive to $\sigma$ (Figure~\ref{fig:snd3_weight}B), confirming that the weight distribution robustly identifies the same rigid core regardless of regularization strength. 

The role of $\sigma$ is not to alter which residues are mobile, but to control the sharpness of the core-periphery boundary used for alignment and domain decomposition. Larger $\sigma$ yields a diffuse weighting that spreads weight across many residues, while smaller $\sigma$ sharpens the boundary, concentrating weight on the most rigid atoms. However, excessively small $\sigma$ over-regularizes the distribution, collapsing the effective alignment set to too few atoms and risking sensitivity to local structural noise rather than global domain motion.

\begin{figure}[ht!]
  \centering
  \includegraphics[width=0.75\linewidth]{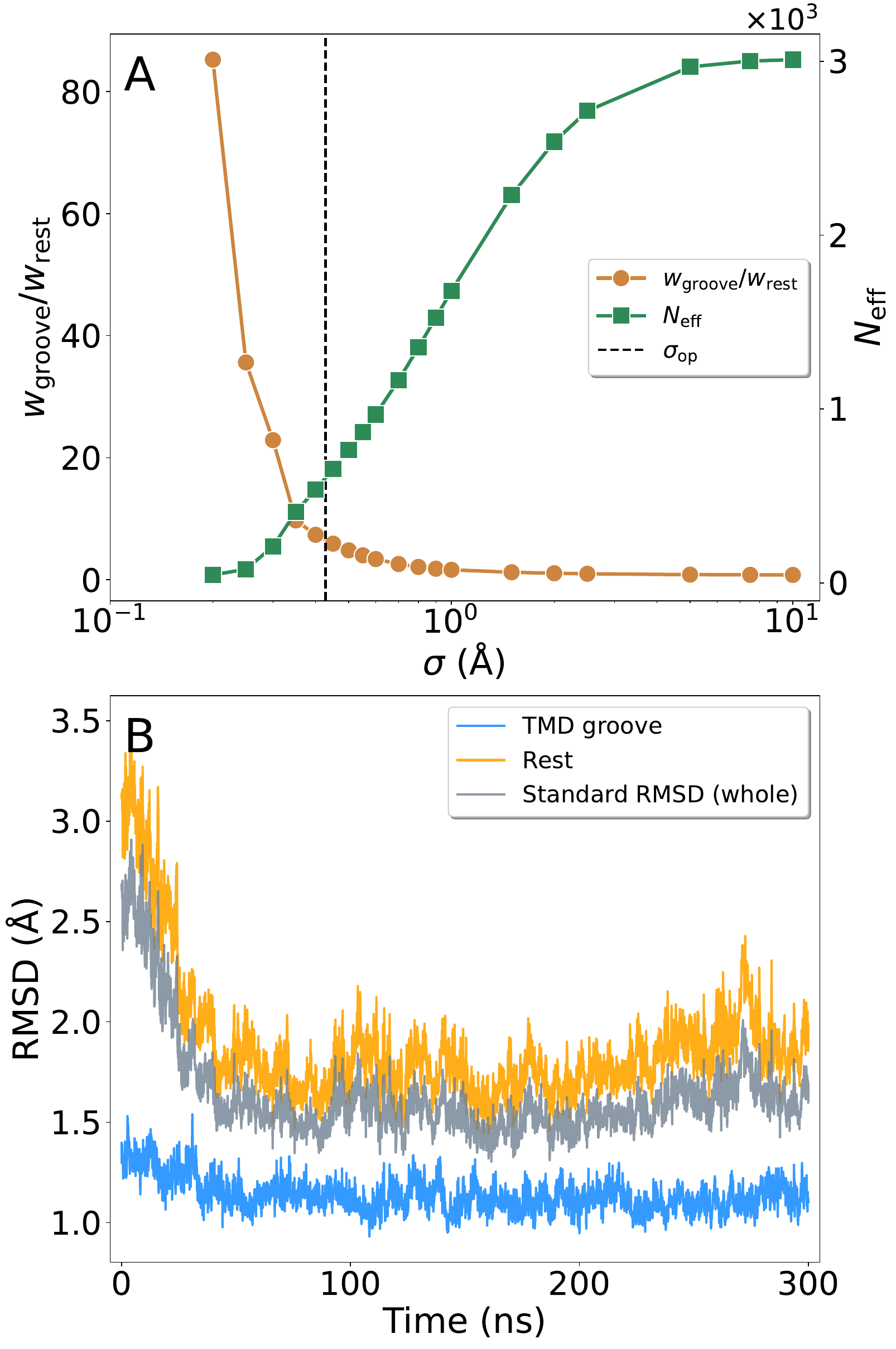}
  \caption{Optimal $\sigma$ selection and alignment quality are shown in two panels. (A) Weight ratio $w_\text{groove}/w_\text{rest}$ (red, left axis) and effective atom count $\neff$ (green, right axis) are plotted versus $\sigma$ (log scale). At $\sigop = 0.43$~\AA{} (dashed line), the ratio is 6.5 and $\neff = 604$. (B) RMSD time series over the 300~ns trajectory for the TM groove (blue), the non-groove region (orange), and overall standard RMSD (gray). The groove RMSD is substantially lower and less variable than the others.}
  \label{fig:snd3_opt}
\end{figure}

The optimal $\sigma$ is chosen from two diagnostics plotted in Figure~\ref{fig:snd3_opt}A, namely the ratio $w_\text{groove}/w_\text{rest}$ with $w_\text{D} = \sum_{\alpha\in{\rm D}}w_\alpha\;(D={\rm groove, rest})$ (shown in red) and the effective atom count $\neff$ (eq~\ref{eq:neff}) (shown in green). We define $\sigop$ as the smallest $\sigma$ at which $\neff \geq 0.20\,N$. At this value, at least one-fifth of all atoms contribute to the alignment and weight concentration on the rigid core is near its practical maximum. At $\sigop = 0.43$~\AA{}, the ratio reaches 6.5 and $\neff = 604$, so the groove residues carry the dominant weight while a statistically meaningful fraction of the chain (604 of 3013 atoms) still contributes to the alignment. The cumulative weight-MSF Lorenz curve at $\sigop$ quantifies this discrimination. The 50\% most-rigid atoms carry approximately 90\% of the total weight (Figure~\ref{figS:lorenz}).

The weight optimization automatically concentrates weight on the TM groove without manual residue selection. The resulting BRMSD ($0.50\pm0.08$~\AA) at $\sigop=0.43$~\AA{} is $\sim$3 times smaller than the standard RMSD ($1.64\pm0.25$~\AA, uniform weights) (Figure~\ref{figS:rmsd_ts}). The domain-decomposed RMSD time series at $\sigop$ (Figure~\ref{fig:snd3_opt}B) shows that the TM-groove RMSD is substantially lower and less variable than both the non-groove domain (Rest) and the standard all-atom RMSD, confirming that weight concentration on rigid atoms suppresses the dominant source of conformational noise. Per-residue RMSF and atom weights at $\sigop$, shown in Figure~\ref{figS:rmsf_weight}, demonstrate the exponential suppression of flexible domains. The BRMSD-aligned trajectory is shown in Movie~S1 alongside the raw trajectory.

\subsection{Focused Alignment}

To demonstrate the focused alignment module, we selected TMD-1 in SND3. Although it is membrane-embedded and structurally ordered, it contributes marginally to the global alignment ($w_\text{TMD1} = 0.002$ at $\sigop$) because its fluctuations are larger than those of TMD-2/3/4. Focused alignment recovers the internal dynamics of TMD-1 by explicitly biasing the weight optimization toward the TMD-1 atoms.

\begin{figure}[ht!]
  \centering
  \includegraphics[width=\linewidth]{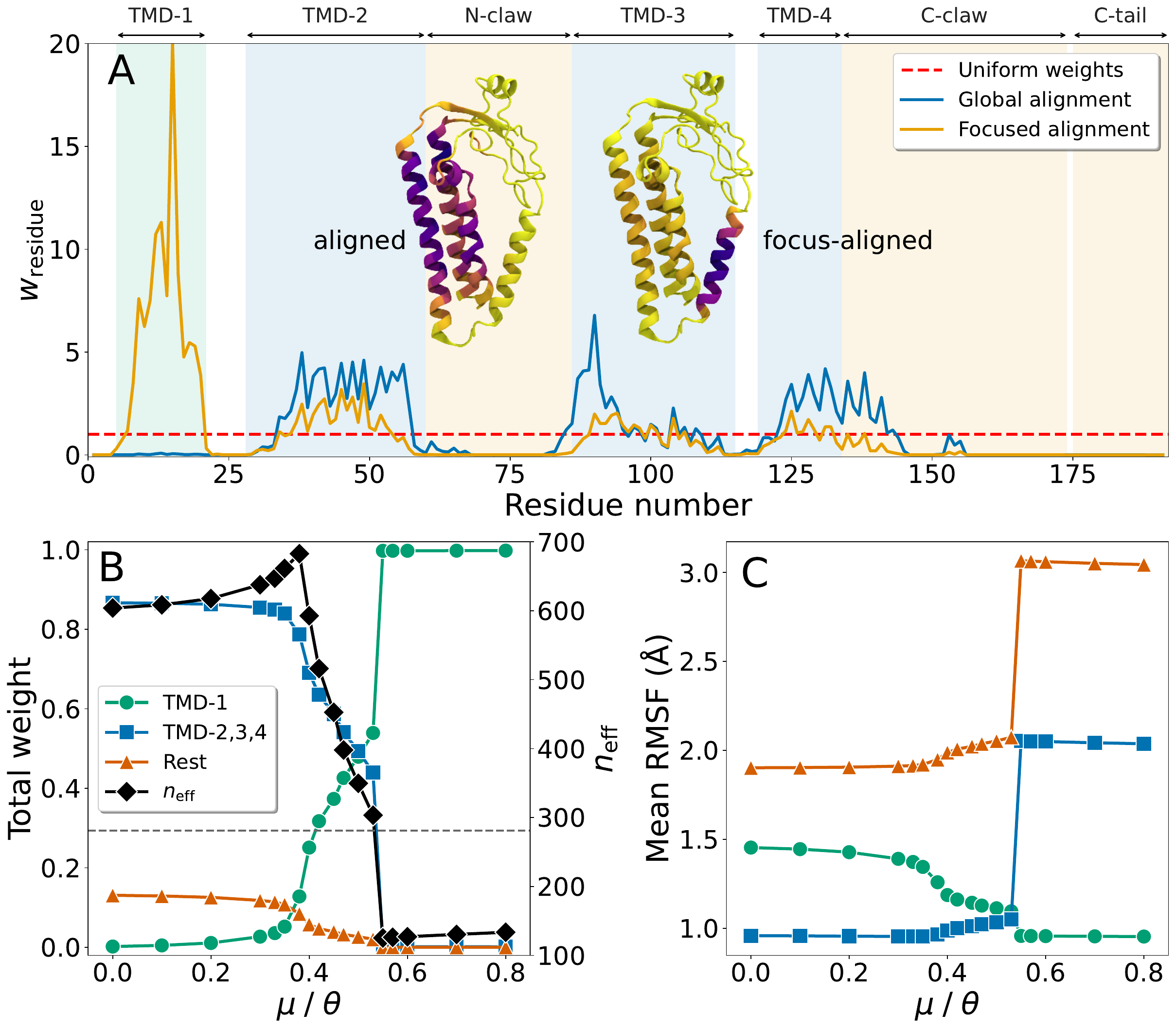}
  \caption{Results of focused alignment of SND3 onto TMD-1 are shown in three panels. (A) Per-residue weights are plotted against residue numbers. Global alignment at $\sigop = 0.43$~\AA{} (blue) and focused alignment at $(\mu/\theta)_\text{op} = 0.53$ (orange) are shown simultaneously along with the uniform baseline (red dashed line). The average structures from each alignment are colored by the weights on each atom, with darker shades denoting higher weights. (B) Total weight fractions on TMD-1 (green), TM groove (blue), and the rest (orange) are plotted as functions of $\mu/\theta$. The effective atom count $\neff$ (black, right axis) and the TMD-1 size $\lvert D\rvert = 281$ (dashed) are overlaid. (C) Mean RMSF of each segment from the focus-aligned average structure is plotted against $\mu/\theta$.}
  \label{fig:focus}
\end{figure}

We scanned $\mu/\theta$ ratio from 0 to 0.8 at $\sigma = 0.43$~\AA{} (Figure~\ref{figS:focus_profiles}). Figure~\ref{fig:focus}A shows the per-residue normalized weight profiles for global alignment at $\sigop$ (blue) and focused alignment at $\mu/\theta = 0.53$ (orange). The latter raises the per-residue weights sharply in TMD-1 while the TM-groove weight decreases and the non-groove rest remains suppressed. As $\mu/\theta$ increases, $w_\text{TMD1}$ fraction grows from 0.002 to 0.54 while the $w_\text{groove}$ fraction decreases from 0.87 to 0.44 and $w_\text{rest}$ fraction falls to 0.02 (Figure~\ref{fig:focus}B). The TMD-1 mean RMSF decreases from 1.49~\AA{} at global alignment ($\mu/\theta = 0$) to 1.14~\AA{} with focused alignment at $\mu/\theta = 0.53$ (Figure~\ref{fig:focus}C). This 24\% reduction reflects the removal of rigid-body translations and rotations of TMD-1 relative to the rest of the protein. The per-residue weight and RMSF profiles for all $\mu/\theta$ values are shown in Figure~\ref{figS:focus_profiles}A.

The operating point $(\mu/\theta)_\text{op} = 0.53$ is identified from the $\neff$ criterion. At this value $\neff = 303$, which just exceeds the TMD-1 atom count $\lvert D \rvert = 281$ (Figure~\ref{fig:focus}B). At $\mu/\theta = 0.55$, $\neff$ drops below $\lvert D \rvert$ (to 125), marking the onset of numerical over-concentration. Beyond the crossover at $\mu/\theta \approx 0.53$, the focusing penalty overwhelms the entropy term and almost all weight collapses onto TMD-1 ($w_\text{TMD1}\to1.0$, $w_\text{groove}\to0.002$; Figure~\ref{fig:focus}B). The alignment is then performed on TMD-1 alone, so the now-unweighted TM groove and rest of the protein are no longer superimposed and their RMSF rises sharply (Figure~\ref{fig:focus}C). This over-concentration, where $\neff<\lvert D\rvert$, marks the upper usable limit of $\mu/\theta$. The per-$\mu/\theta$ RMSF profiles in Figure~\ref{figS:focus_profiles}B confirm that the groove and rest RMSF values are largely insensitive to the focusing parameter, isolating the TMD-1 RMSF reduction as a genuine alignment effect rather than an artifact.

\subsection{Trajectory Smoothing}

We smoothed the SND3 trajectory using a triangular window kernel, sweeping the half-width $W$ over 5, 10, 20, 30, and 50 frames (full window 1--10.0~ns) at $\sigop = 0.43$~\AA{} (Figure~\ref{figS:smooth_sweep}). We quantify the smoothing quality by the mean BRMSD of each raw frame from its locally windowed average. This quantity increases with window width because wider windows produce local averages further from any individual frame, ranging from $0.30$~\AA{} at $W = 5$ frames to $0.37$~\AA{} at $W = 50$ frames. We adopt $W_\text{op} = 20$ frames (4.0~ns full window). The 4.0~ns full window also sits below the slow TM-groove modulation timescale, so it suppresses sub-nanosecond fluctuations without averaging out the slower collective motion. The mean BRMSD from the windowed average at $W_\text{op}$ is $0.35$~\AA{}.

At $W_\text{op}$, the mean BRMSD of the smoothed trajectory from the global average drops from $0.50 \pm 0.08$~\AA{} (raw) to $0.35 \pm 0.04$~\AA{}, a 30\% reduction (Figure~\ref{figS:smooth_ts}). The smoothed trajectory retains the slow TM groove modulations visible at timescales above 4~ns while suppressing sub-nanosecond fluctuations. Movie~S2 shows the raw and smoothed trajectories side by side. Per-residue RMSF in Figure~\ref{figS:smooth_rmsf} shows that smoothing reduces fast fluctuations uniformly across the chain, while the slow, large-amplitude movements are retained.

\subsection{Conformational Clustering}

Unlike alignment and domain identification, conformational clustering of the open-to-closed transition requires a broad weight distribution rather than a concentrated one. The signal that separates the two states of ADK resides in the mobile NMPbind and LID domains, so concentrating weight on the rigid CORE (small $\sigma$) suppresses precisely those displacements and collapses the cluster centres onto a single point. A sweep of the alignment $\sigma$ shows that the Calinski--Harabasz (CH) index (eq~\ref{eq:ch}) is almost zero for $\sigma \lesssim 2$~\AA{}, where the cluster centers merge. CH index rises sharply between $\sigma=3$ and $4$~\AA{} and plateaus for larger $\sigma$. We adopt $\sigma=4.0$~\AA{} as the onset of this stable, high-CH regime (Figure~\ref{figS:cluster_sigma}).

To select $\tau$, we swept over the values 1, 2, 5, 10, 20, and 50~\AA$^2$ at $K = 2$, evaluating the CH index (eq~\ref{eq:ch}), minimum centre separation (eq~\ref{eq:dmin}), and population balance (Figure~\ref{figS:tau_sweep}). The CH index and centre separation remain high for $\tau \le 5$~\AA$^2$ and collapse to zero at $\tau \ge 10$~\AA$^2$, where the responsibilities become uniform and the two centres merge. We adopt the largest $\tau$ that preserves discrimination. At $\taup = 5.0$~\AA$^2$ the two centres are well separated (centre separation $= 3.20$~\AA{}) and the partition is most balanced ($|\bar{q}_1 - 0.5| = 0.02$), whereas smaller $\tau$ over-sharpens toward hard $K$-means without improving the partition.

$K$-screening at $\taup$ identifies $K = 2$ as optimal. At $K = 3$, the minimum inter-centre separation collapses to 0.0~\AA{} across restarts, indicating fully degenerate centres. No third distinct conformational state was detectable by BRMSD clustering at these parameters.

\begin{figure}[ht!]
  \centering
  \includegraphics[width=\linewidth]{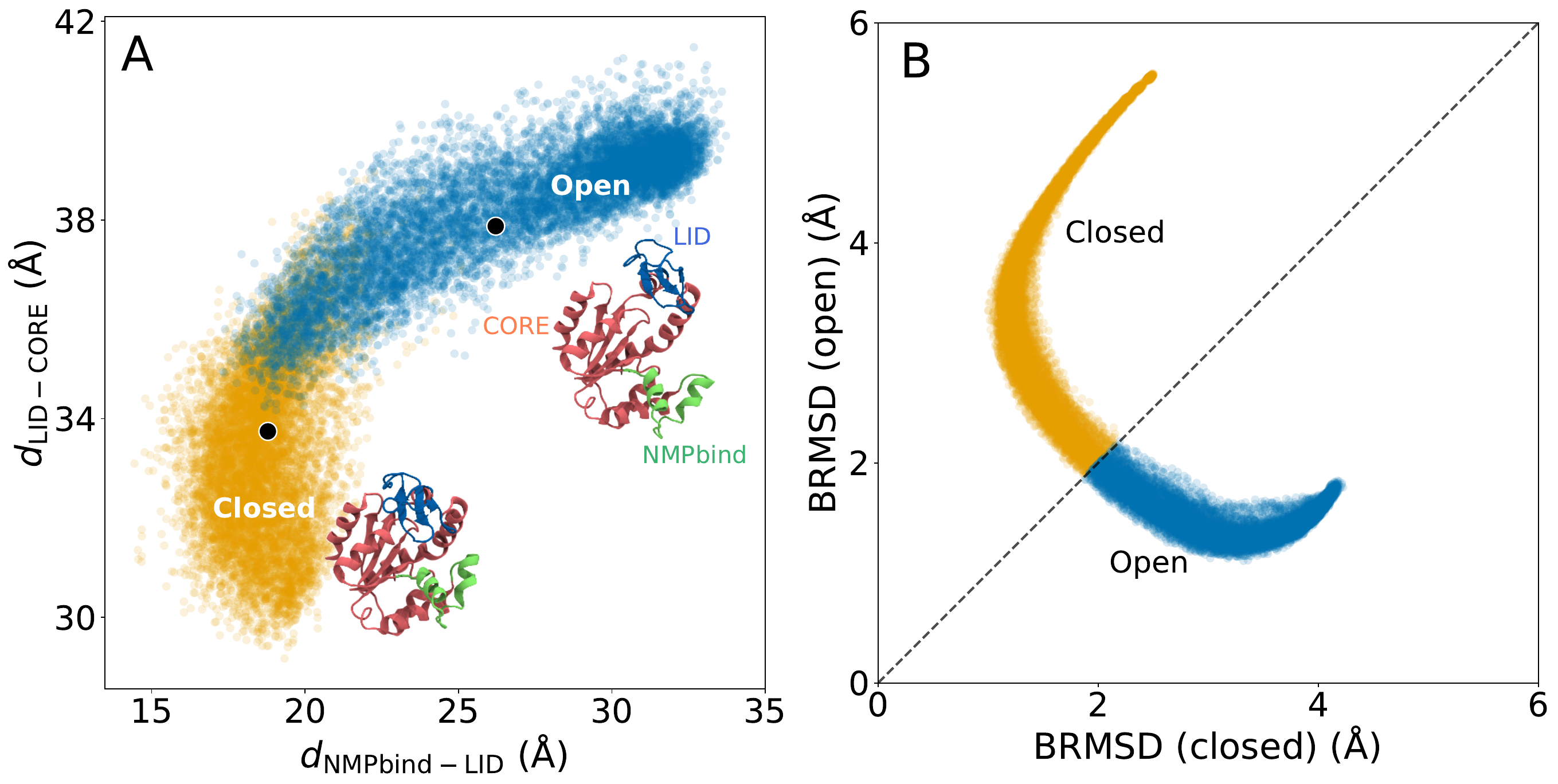}
  \caption{Conformational clustering of the 19,691-frame ADK DIMS ensemble ($K = 2$, $\taup = 5.0$~\AA$^2$, $\sigma = 4.0$~\AA{}). (A) Each frame placed by its order parameters $d_{\rm NMPbind-LID}$ and $d_{\rm LID-CORE}$, colored by hard cluster assignment: orange (closed) and blue (open). Black circles mark the two cluster centers, shown colored by domain (CORE red, LID blue, NMPbind green). (B) The same frames placed by their weighted BRMSD to the closed center (abscissa) and to the open center (ordinate), each computed with that cluster's own weight vector. The dashed line marks equal BRMSD from both clusters.}
  \label{fig:adk_cluster}
\end{figure}

The two clusters correspond to open-like (soft population 51.8\%, hard count 10,610 frames) and closed-like (soft population 48.2\%, hard count 9,081 frames) conformations (Figure~\ref{fig:adk_cluster}). The corresponding cluster centres (responsibility-weighted average structures, eq~\ref{eq:sa_update}) are located at $d_{\rm NMP-LID}$ = 26.2~\AA{}, $d_{\rm LID-CORE}$ = 37.9~\AA{} (open) and $d_{\rm NMP-LID}$ = 18.8~\AA{}, $d_{\rm LID-CORE}$ = 33.7~\AA{} (closed). They are denoted by the black circles in Figure~\ref{fig:adk_cluster}A. The near equal populations reflect the pathway-sampling design of the DIMS ensemble and do not correspond to thermodynamic equilibrium free energies.\cite{perilla2011} DIMS trajectories sample structural coverage of the open-to-closed transition, not a Boltzmann distribution, so population fractions cannot be interpreted as equilibrium state probabilities.

Plotting each frame by its weighted BRMSD to the two cluster centres (Figure~\ref{fig:adk_cluster}B), each distance evaluated in that cluster's own weight vector, separates the ensemble into two off-diagonal arcs that meet only along the $d_\text{closed} = d_\text{open}$ diagonal. Because every axis uses a cluster-adapted metric, a frame of one state measured against the other centre is driven to large distance by precisely the mobile LID and NMPbind atoms that distinguish the two states, giving a cleaner separation than the geometric order parameters of Figure~\ref{fig:adk_cluster}A. The diagonal crossing marks the transition dividing surface ($q \approx 0.5$), and its low point density indicates rapid transit through the barrier region. The outward curvature of each arc reflects the arced, hinge-bending transition path between the open and closed endpoints rather than a straight interpolation between the two centres.

The responsibilities ($q(a|i)$, where $a$ is the cluster and $i$ is the structure) correlate strongly with both order parameters. Pearson correlation coefficient of $q(\text{cluster 1}|i)$ with $d_\text{NMP-LID}$ and $d_\text{LID-CORE}$ are $-0.88$ and $-0.94$, respectively (Figure~\ref{figS:responsibilities}). The median of the ratio $d_\text{between}/d_\text{within}$ (between-cluster and within-cluster distances) is 2.48 and all frames have $d_\text{between} > d_\text{within}$. This confirms that the two conformational states are genuinely distinct in BRMSD space (Figure~\ref{figS:separation}). Per-residue backbone displacement between the two cluster centres (Figure~\ref{figS:cluster_displacement}) reveals the hierarchical displacement order. LID (mean 5.68 \AA) $>$ NMPbind (4.73 \AA) $\gg$ CORE (1.11 \AA), consistent with the ground-truth domain assignments.

\subsection{Rigid Domain Identification}

We first applied the domain identification module to SND3 to assess whether any sub-domain structure is resolvable. Sequential peeling recovers the domains when the alignment $\sigma$ is set to the rigidity scale of TMD-1. A $\sigma$ sweep locates the cleanest separation at $\sigma = 1.0$~\AA{} (Figure~\ref{figS:snd3_peeling_selection}), where round~1 recovers the TM groove and round~3 recovers TMD-1 as its own domain (Figure~\ref{figS:snd3_peeling_weights}) with a Jaccard index of 0.44 to the TMD-1 reference (Figure~\ref{figS:snd3_peeling_jaccard}). Round 2 falls on the cytosolic claws but does not form a clean domain ($J = 0.28$). The separation is sharp in $\sigma$. It peaks at 1.0~\AA{} and collapses for $\sigma \geq 1.6$~\AA{}, where TMD-1 merges into the groove. This value matches the internal RMSF of TMD-1 measured by focused alignment (1.14~\AA{}). Only the stiffest half of TMD-1 is recovered in round 3 because it is a partially rigid membrane-embedded helix rather than a fully independent domain.

For ADK, we applied sequential peeling to the 28,282-frame FRODA backbone ensemble. FRODA generates motions by propagating rigid-body constraints on covalent geometry, so its domain boundaries are structurally unambiguous by construction. Hence the domains are recovered more cleanly in ADK than SND3 (300 ns equilibrium run). However, simultaneous $K = 3$ optimization fails to detect the three different rigid domains in ADK. This happens because all three weight vectors collapse onto CORE regardless of the overlap penalty $\mu$. Every domain begins with uniform weights, so the first iteration concentrates all vectors onto the globally lowest-MSF region (CORE), from which the overlap-penalty gradient is too weak to escape. Sequential peeling resolves this by running three independent $K = 1$ optimizations on shrinking atom pools. The peeling order follows global rigidity. The most rigid domain is always found first, so the order (CORE $\to$ LID $\to$ NMPbind) is data-determined rather than user-specified.

\begin{figure}[ht!]
  \centering
  \includegraphics[width=\linewidth]{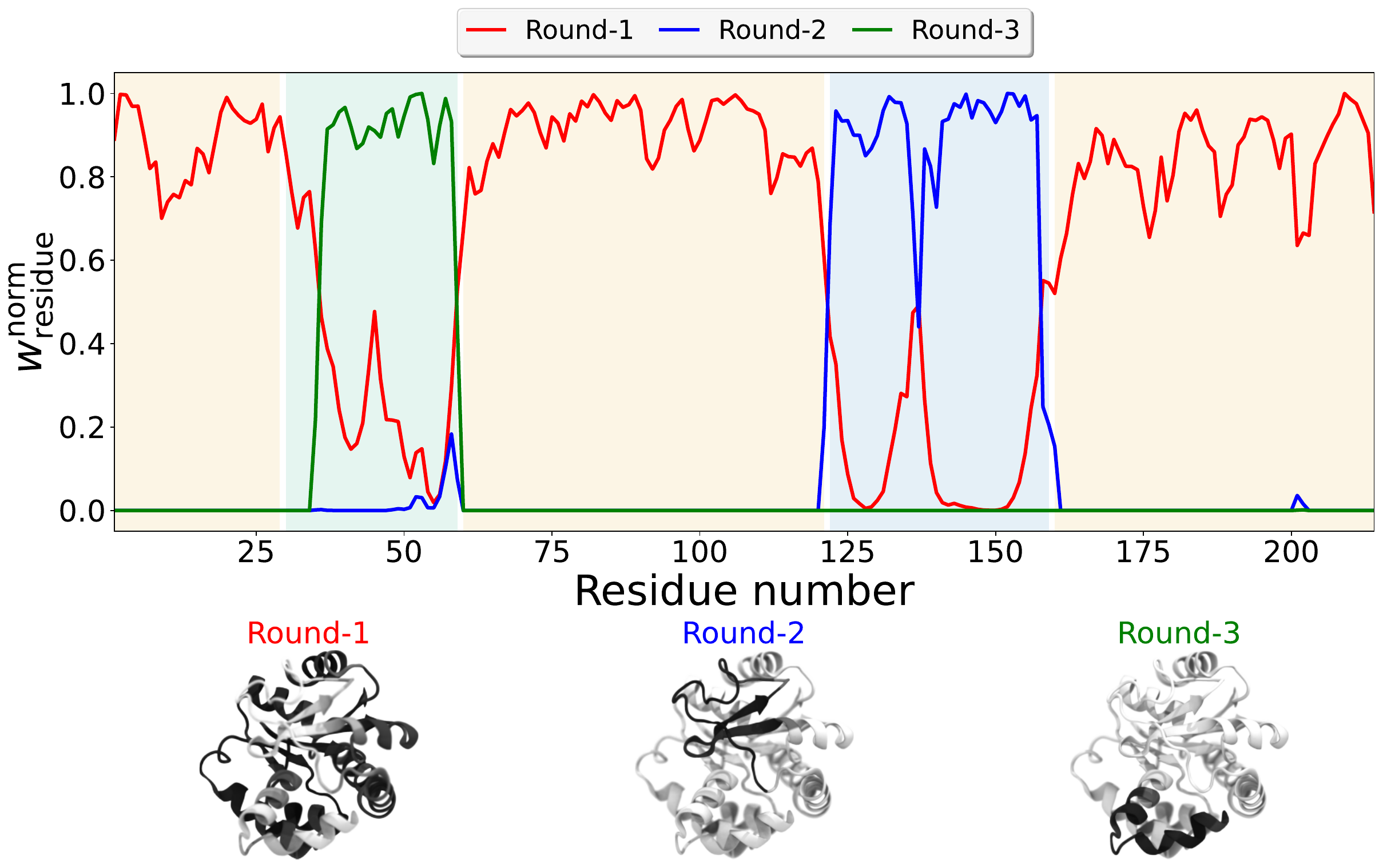}
  \caption{Rigid domain identification for ADK by sequential peeling on 28,282 FRODA backbone frames (856 atoms, $\sigma_\text{op} = 3.0$~\AA{}). The per-residue weight profiles $w^{\rm norm}_{\rm residue}$ (mean atom weight of a residue, normalized within each round to its maximum) for the three peeling steps are overlaid: Round-1 (red, CORE), Round-2 (blue, LID), and Round-3 (green, NMPbind), with domain shading. The corresponding average structures, colored by weight, are shown below. Domains are detected in order of decreasing global rigidity. }
  \label{fig:adk_domain}
\end{figure}

 A $\sigma$ sweep at $K = 1$ on the full 856-atom pool identifies $\sigma_\text{op} = 3.0$~\AA{} as maximizing $J_\text{CORE}$ (Figure~\ref{figS:adk_sigma_sweep}). Here $J_X$ is the Jaccard score for domain $X$, which measures the overlap between the atoms recovered by the BRMSD module and the experimental ground truth, so this criterion requires ground-truth domain labels. Figure~\ref{fig:adk_domain} reports the recovered domains through the per-residue weight (the mean BRMSD weight of each residue's backbone atoms). Each round is normalized by its own maximum, so that the most rigid residue in a round equals one. Round~1 (856 atoms, 5 restarts) recovers CORE with $J_\text{CORE} = 0.94$ and claims 616 atoms above the $t = 0.50$ threshold. Round~2 (240 non-CORE atoms, 5 restarts) identifies the LID domain ($J_\text{LID} = 0.99$), and Round~3 (96 remaining atoms, 5 restarts) recovers NMPbind with $J_\text{NMPbind} = 1.00$. The Jaccard scores for all three domains are given in Figure~\ref{figS:adk_jaccard}. At $t = 0.50$ the CORE mask absorbs 26 of the 120 NMPbind backbone atoms at the CORE--NMPbind boundary, but the remaining NMPbind and LID cores stay cleanly separated, so both flexible domains are recovered essentially perfectly.

A sensitivity sweep over $t \in [0.40, 0.95]$ shows that $J_\text{CORE}$ is flat (the step-1 optimization is independent of $t$), while $J_\text{LID}$ and $J_\text{NMPbind}$ degrade monotonically as $t$ increases. For $t \leq 0.60$, the CORE mask slightly overclaims ($N_\text{core} = 580$--636 vs.\ the ground-truth 584), but the non-CORE pool still separates cleanly into LID and NMPbind, giving $J_\text{LID} \geq 0.94$ and $J_\text{NMPbind} = 1.00$. As $t$ increases beyond 0.70, CORE is progressively underclaimed ($N_\text{core} = 413$ at $t = 0.80$, 201 at $t = 0.90$), leaving rigid CORE hinge residues in the pool that dominate the later steps and collapse both $J_\text{LID}$ and $J_\text{NMPbind}$ toward zero.

\subsection{Consistency and Uniqueness of the Optimal Weights}

Under the $\theta = M\sigma^2$ convention, each atom's converged weight depends only on its own mean-square fluctuation relative to $\sigma$, independent of the number of frames $M$ and of the number of atoms $N$ in the selection. We verified both invariances on the SND3 alignment at $\sigop = 0.43$~\AA{} (Figure~\ref{figS:consistency}). We quantify the difference between two normalized weight vectors $\boldsymbol{w}$ and $\boldsymbol{w}'$ by the Jensen--Shannon distance, the square root of the Jensen--Shannon divergence (a symmetrized form of the KL divergence).
\begin{equation}
    D_{\rm JS}(\boldsymbol{w}|\boldsymbol{w}') = \left[\frac{1}{2\ln 2}\sum_\alpha\left(w_\alpha\ln\frac{w_\alpha}{m_\alpha} + w'_\alpha\ln\frac{w'_\alpha}{m_\alpha}\right)\right]^{1/2},
    \qquad m_\alpha = \tfrac{1}{2}\left(w_\alpha + w'_\alpha\right),
    \label{eq:djs}
\end{equation}
$D_{\rm JS}$ is bounded in $[0,1]$ and is well defined even when some weights vanish. $D_{\rm JS} = 0$ for identical weight vectors and $D_{\rm JS} = 1$ for distributions with disjoint support.

Recomputing the weights from alternate frames (1501 of 3001 frames) leaves the per-residue weights unchanged ($D_\text{JS} = 0.012$), confirming $M$-invariance (Figure~\ref{figS:consistency}A). Removing an entire mobile region from the selection and re-optimizing on the retained atoms likewise leaves their weights unchanged. Dropping TMD-1 ($D_\text{JS} = 0.014$) (Figure~\ref{figS:consistency}B) or the cytosolic claws ($D_\text{JS} = 0.074$) (Figure~\ref{figS:consistency}D) has almost no effect, whereas removing the rigid TM groove, which carries the dominant weight, forces a different solution, yielding a higher $D_\text{JS} = 0.827$ (Figure~\ref{figS:consistency}C). The weights are thus invariant to random changes in the ensemble but responsive to targeted changes.

The weight update is a nonlinear fixed-point map, so the free energy $G$ can in principle support several self-consistent solutions. Seeding the SND3 alignment from 40 spatially localized weight vectors (Gaussian bumps of half-width 10~\AA{} centered on random atoms) and clustering the converged solutions by $D_\text{JS}$ shows that, at the operating point $\sigop$, all initializations converge to a single solution (Figure~\ref{figS:minima}). As $\sigma$ decreases below $\sigop$, the landscape fragments into multiple competing minima (Figure~\ref{figS:minima}A), each concentrating weight on a different local rigid sub-structure (Figure~\ref{figS:minima}C), and all with higher $G$ than the global minimum (Figure~\ref{figS:minima}B). This justifies the single initialization used for the alignment and smoothing modules at $\sigop$. It also explains why clustering and domain identification use random restarts.

\section{Conclusions}

BRMSD unifies multiple structural analysis tasks within a single Bayesian framework. This includes global alignment, focused alignment, trajectory smoothing, conformational clustering, and rigid domain identification. All five use a free-energy functional with a concentration parameter $\sigma$ (via $\theta = M\sigma^2$), which establishes the MSF threshold such that atoms with fluctuations well below the $\sigma$ scale are assigned near-uniform weights, effectively preventing the metric from focusing on local noise. The optimal $\sigma$ is module- and system-specific. Parameter selection follows the same diagnostic in each module, monitoring $\neff$ or a module-specific quality index as a function of $\sigma$. Because the weights are set by the per-atom MSF, they are insensitive to trajectory length once the MSF estimates have converged. For short trajectories where the per-atom MSF has not yet converged, $\sigop$ and the recovered weight profiles should be treated as provisional.

BRMSD was benchmarked against two well-resolved systems. In SND3, alignment at $\sigop = 0.43$~\AA{} reduces the apparent RMSD threefold relative to standard backbone superposition, automatically suppressing mobile cytosolic claws without manual residue selection. Focused alignment onto TMD-1 at $(\mu/\theta)_\text{op} = 0.53$ reduces the TMD-1 RMSF by a further 24\%, revealing its internal dynamics against the background of TM groove motion. Smoothing with a 4~ns triangular window reduces BRMSD variance by $\sim$73\% while preserving slow TM groove modulations. 

In ADK, soft $K$-means clustering with cluster-specific weight vectors recovers the known open/closed bimodality. The cluster centres (responsibility-weighted average structures) align with the expected order-parameter regions of the conformational landscape. Sequential domain detection recovers all three known ADK domains in order of decreasing global rigidity (CORE, LID, and NMPbind). For SND3, simultaneous optimization finds no sub-domain partition, and sequential peeling at $\sigma = 1.0$~\AA{} partially recovers TMD-1 as a separate rigid unit, at the same fluctuation scale identified by focused alignment.

The contribution of BRMSD is to automate, reproducibly and within a single framework, the atom weighting that one would otherwise assign manually. A small $\sigma$ recovers a rigid core without manual selection, whereas a large $\sigma$ recovers the standard-RMSD metric where the discriminating signal is distributed. However, in its current form \texttt{BRMSD} has certain limitations, as listed below. 
\begin{enumerate}
    \item The operating point $\sigma$ is task-dependent and is not governed by a single universal rule. The converged weights depend on per-atom MSF convergence. So for short trajectories $\sigop$ and the weight profiles are provisional.

    \item For conformational clustering and rigid domain identification, the operating parameters $\sigma_\text{op}$, $\taup$, and the peeling threshold $t$ were selected using quality indices (CH and Jaccard) that reference the known cluster or domain assignments. A fully label-free selector for these parameters remains an open problem.

    \item The current implementation of \texttt{BRMSD} is CPU-only. 
    
    \item The Mahalanobis-like extension for anisotropic fluctuations (eq~\ref{eq:GPCA}) is derived but not yet implemented.
\end{enumerate}

The current implementation uses a self-consistent fixed-point iteration to find the entropy-optimal weights. The converged BRMSD weights could serve as an atom-selection mask for time-lagged independent component analysis (TICA)\cite{noe2015,schultze2021}. Restricting the coordinate input to atoms with $w_\alpha$ above a threshold removes noise-dominated degrees of freedom before slow-mode identification, which should improve MSM-based kinetic models.\cite{husic2018} 

Training a neural network to predict per-atom weights from local structural features, using the maximum-entropy loss $G$ as the training objective, would bypass the fixed-point iteration and make the method applicable to large ensembles at low marginal cost. Related work has combined neural networks with maximum entropy principles for active learning of conformational ensembles,\cite{kleiman2023} where maximum entropy guides the choice of which conformations to simulate next. Here we instead propose using $G$ as a training loss for predicting what weights to assign, a complementary application of the same principle. Further directions include GPU-accelerated kernels for large ensembles ($M\gtrsim10^5$ or $N\gtrsim10^4$), implementing the Mahalanobis-like PCA-mode weighting of eq~\ref{eq:GPCA}, and a quantitative, label-free $\sigma$ selector for clustering and domain detection.

\begin{acknowledgement}
S.M. thanks the Max Planck Society (MPG) for research funding and the Max Planck Computing and Data Facility (MPCDF) for computational support. G.H. acknowledges the German Research Foundation (DFG) support through SFB 1507/P12.
\end{acknowledgement}

\section*{Author Contributions}
S.M. designed and implemented the BRMSD package, performed all benchmark computations, and drafted the manuscript. G.H. conceived the Bayesian framework, supervised the work, and revised the manuscript.

\section*{Data and Code Availability}
The \texttt{BRMSD} package is openly available at \url{https://github.com/bio-phys/BRMSD} under the MIT license. The MD simulation files of SND3 are available on Zenodo (\href{https://doi.org/10.5281/zenodo.16745188}{16745188})\cite{Zenodo2025}. The ADK trajectories (DIMS and FRODA) are distributed through the MDAnalysisData repository\cite{mdanalysisdata}.

\clearpage
\setcounter{figure}{0}
\renewcommand{\thefigure}{S\arabic{figure}}
\setcounter{table}{0}
\renewcommand{\thetable}{S\arabic{table}}
\setcounter{equation}{0}
\renewcommand{\theequation}{S\arabic{equation}}

\section*{Supporting Information}

\subsection*{Clustering and Domain Detection Quality Metrics}

\subsubsection*{Calinski--Harabasz (CH) Index}
The CH index~\cite{Calinski1974} measures the ratio of between-cluster to within-cluster scatter.
\begin{equation}
\text{CH}(K) = \frac{B(K)/(K-1)}{W(K)/(M-K)},
\label{eq:ch}
\end{equation}
where $M$ is the number of frames, $B(K)$ is the total between-cluster scatter (sum of cluster-size-weighted squared distances from $K^{th}$ cluster centroid to the global centroid), and $W(K)$ is the total within-cluster scatter (sum of squared distances from each frame to its assigned cluster centroid), both computed in BRMSD space. A higher CH indicates clusters that are simultaneously more compact and better separated.

\subsubsection*{Minimum Inter-centre Separation}
The minimum squared Euclidean distance between any two cluster centres in BRMSD space,
\begin{equation}
d_\text{min}^2(K) = \min_{i \neq j}
\MSD(s_i, s_j; \bvec{w}),
\label{eq:dmin}
\end{equation}
detects degenerate solutions in which two centres collapse onto the same point.
At $K = 3$ for the ADK ensemble, $d_\text{min} = 0.0$~\AA{}, confirming that the third centre has no distinct structural identity. This check is necessary because CH can be undefined or misleading when $W(K) \approx 0$ for a degenerate cluster.

\subsubsection*{Population Balance}
For $K = 2$, the scalar $|\bar{q}_1 - 0.5|$ measures the deviation of the cluster-1 population fraction $\bar{q}_1 = M^{-1}\sum_i q(1|i)$ from an equal split, where $q(1|i) \in [0,1]$ is the soft responsibility of frame $i$ for cluster 1. A value near zero indicates a balanced partition, and a value near 0.5 indicates that one cluster absorbs nearly all frames. We report this metric alongside CH because a high-CH solution that is also severely imbalanced ($|\bar{q}_1 - 0.5| \gtrsim 0.3$) may reflect a single dominant state with outliers rather than two physically distinct conformations. At $\taup = 5.0$~\AA$^2$, $|\bar{q}_1 - 0.5| = 0.02$, confirming a near-equal open/closed partition.

\subsubsection*{Jaccard Index}
\begin{equation}
  J(A,B) = \frac{|A \cap B|}{|A \cup B|},
\label{eq:jaccard}
\end{equation}
where $A$ is the set of atoms in the recovered domain (binarized at $w_{k\alpha} > t \cdot \max_\alpha w_{k\alpha}$) and $B$ is the ground-truth domain atom set. $J = 1$ indicates perfect recovery, and $J = 0$ indicates no overlap. We use $t = 0.50$ and report domains with $J > 0.5$ as successfully recovered.

\subsection*{Implementation}

All five modules are implemented in the open-source Python package \texttt{BRMSD}. The package uses NumPy for numerical kernels, MDAnalysis\cite{mda2011,mda2016} for trajectory I/O and atom selection, and joblib for optional multi-threaded parallelism over restarts. Each module exposes a self-contained function (\texttt{BRMSD\_iterate}, \texttt{BRMSD\_focus\_iterate}, \texttt{BRMSD\_smooth}, \texttt{soft\_kmeans}, \texttt{identify\_domains\_sequential}) with a common interface: coordinate array, $\sigma$ value, and module-specific hyperparameters ($W$, $\mu/\theta$, $\tau$, $K$, $\mu$, $n_r$). Python~3.10+ and MDAnalysis~2.x are required.

For the alignment and smoothing modules, a single initialization from the first frame ($s^{(1)} = x_1$, $w_\alpha^{(1)} = W_\alpha$) is used. Convergence is declared when the per-iteration changes in both the average structure and the atom weights fall below the tolerance \texttt{tol} (default $10^{-3}$), that is $\Delta_\text{struct} < \texttt{tol}$ and $\Delta_\text{weights} < \texttt{tol}$. For clustering and domain identification, $n_r$ restarts are run in parallel using joblib, each initialized with random center selection. The restart with the lowest converged $G$ is retained. For the benchmarks reported here, $n_r = 3$ restarts are used for clustering and $n_r = 5$ for domain identification. In both cases, the selected solution was reproduced in at least two of the $n_r$ runs, confirming reproducibility of the reported optima.

The per-iteration computational cost scales as $\mathcal{O}(MN)$ for the weight update and $\mathcal{O}(N^3)$ for the Kabsch rotation, where $M$ is the number of frames and $N$ the number of atoms in the selection. Practical runtimes are summarized in Table~\ref{tab:timing}. Alignment of the SND3 trajectory ($M = 3001$, $N = 3013$ all-atom) converges in 31 iterations. Smoothing adds a multiplicative factor of $M$ over alignment but benefits from multi-process parallelism over frames via joblib. Focused alignment converges similarly to standard alignment at each $\mu/\theta$ value. Clustering and domain detection use joblib parallelism over restarts. Systems with substantially larger frame counts ($M \gtrsim 10^5$) or atom selections ($N \gtrsim 10^4$) would benefit from GPU-accelerated matrix operations, which the current implementation does not support.

\begin{table}
  \caption{Computational timing for all benchmark runs on a single CPU core unless otherwise noted (Python 3.12, AMD EPYC server node).}
  \label{tab:timing}
  \begin{tabular}{llllll}
    \toprule
    Module & System & $M$ & $N$ & Restarts & Wall time \\
    \midrule
    Alignment      & SND3  & 3001  & 3013 & 1 & 52~s \\
    Focused align. & SND3  & 3001  & 3013 & 1 per $\mu/\theta$ & 50~s each \\
    Smoothing      & SND3  & 3001  & 3013 & 1 & 349~s (parallel) \\
    Clustering     & ADK DIMS & 19691 & 3341 & 3 ($\times$3 threads) & 1307~s \\
    Domain (step 1) & ADK FRODA & 28282 & 856 & 5 ($\times$3 threads) & 316~s \\
    Domain (step 2) & ADK FRODA & 28282 & 240 & 5 ($\times$3 threads) & 165~s \\
    Domain (step 3) & ADK FRODA & 28282 & 96 & 5 ($\times$3 threads) & 73~s \\
    \bottomrule
  \end{tabular}
\end{table}

\newpage
\subsection*{Supplementary Figures}

\begin{figure}[ht]
  \centering
  \includegraphics[width=\linewidth]{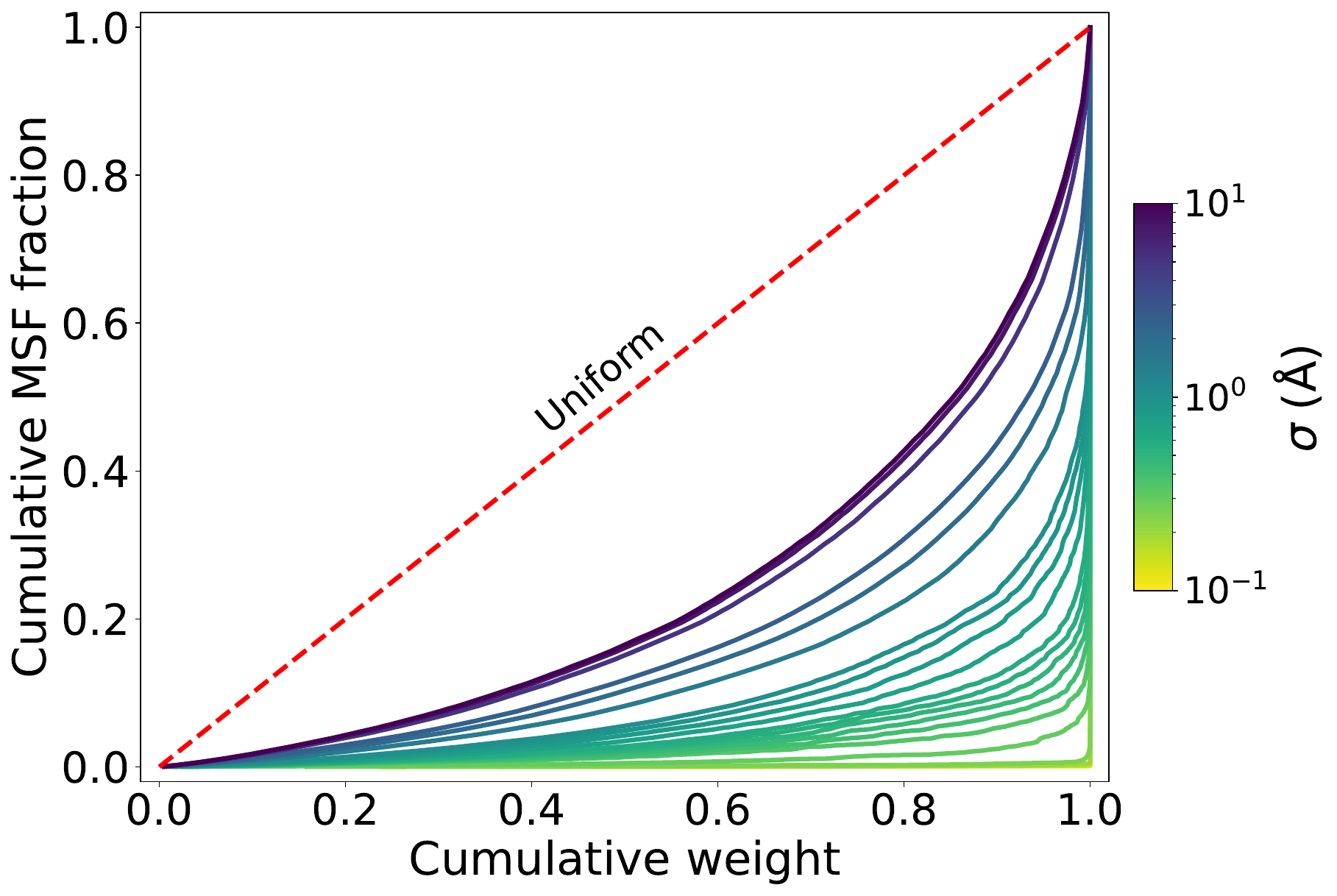}
  \caption{The Lorenz curve (cumulative MSF fraction versus cumulative atom weight) for SND3 is plotted at $\sigma = 0.1$--10.0~\AA{}. As $\sigma$ decreases, the curves depart progressively from the diagonal (equal weights on all atoms), quantifying how weight becomes concentrated on low-MSF (rigid) atoms. At the lowest $\sigma$ values, the most-rigid atoms carry most of the total weight.}
  \label{figS:lorenz}
\end{figure}

\begin{figure}[ht]
  \centering
  \includegraphics[width=\linewidth]{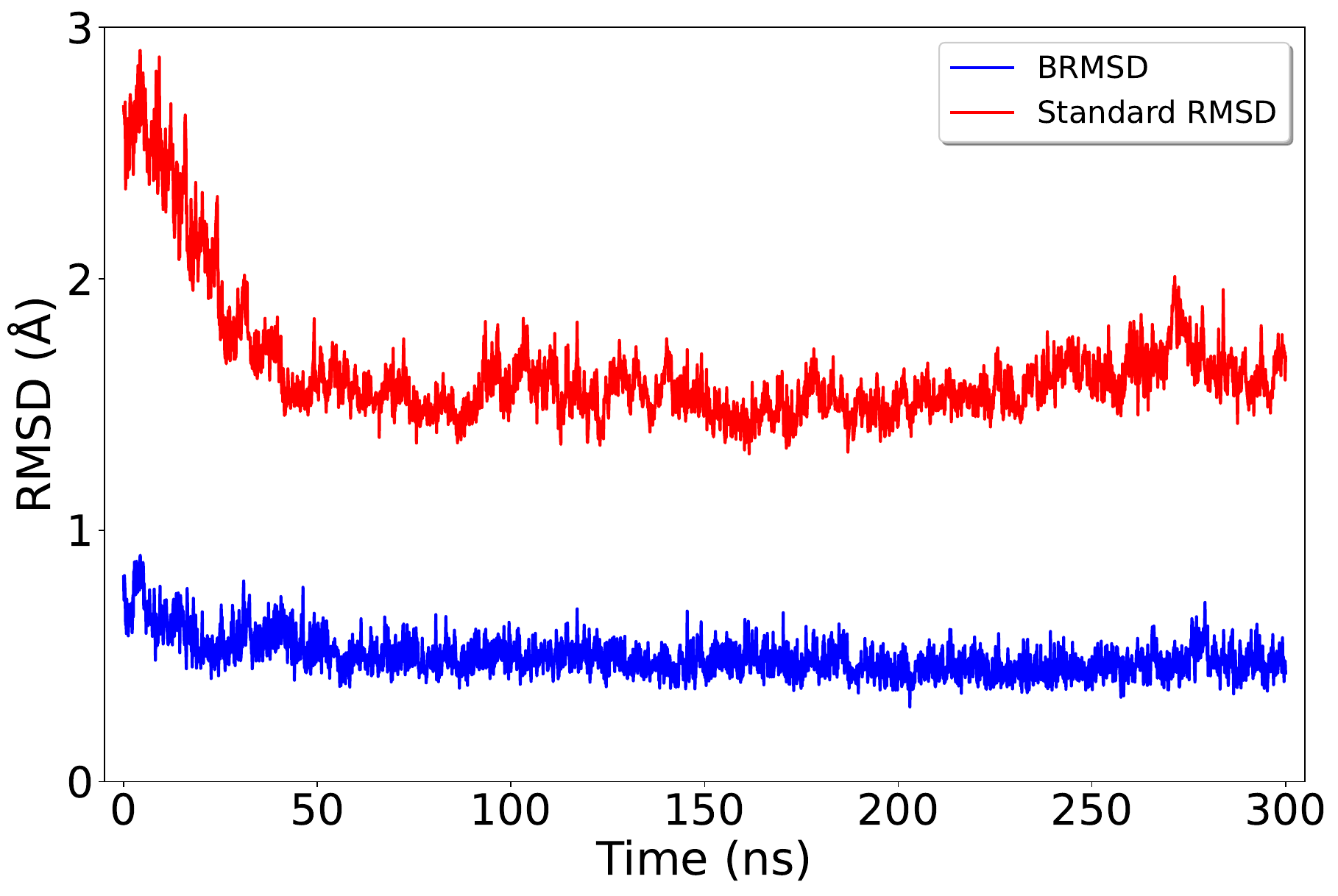}
  \caption{Time series of BRMSD (blue, $\sigma = 0.43$~\AA{}) and standard backbone RMSD (red) for the SND3 300~ns trajectory. The mean BRMSD is $0.50 \pm 0.08$~\AA{} versus a standard RMSD of $1.64 \pm 0.25$~\AA{}, corresponding to a 3.3-fold reduction in the mean and an approximately 10-fold reduction in the variance. Marginal distributions are shown on the right.}
  \label{figS:rmsd_ts}
\end{figure}

\begin{figure}[ht]
  \centering
  \includegraphics[width=\linewidth]{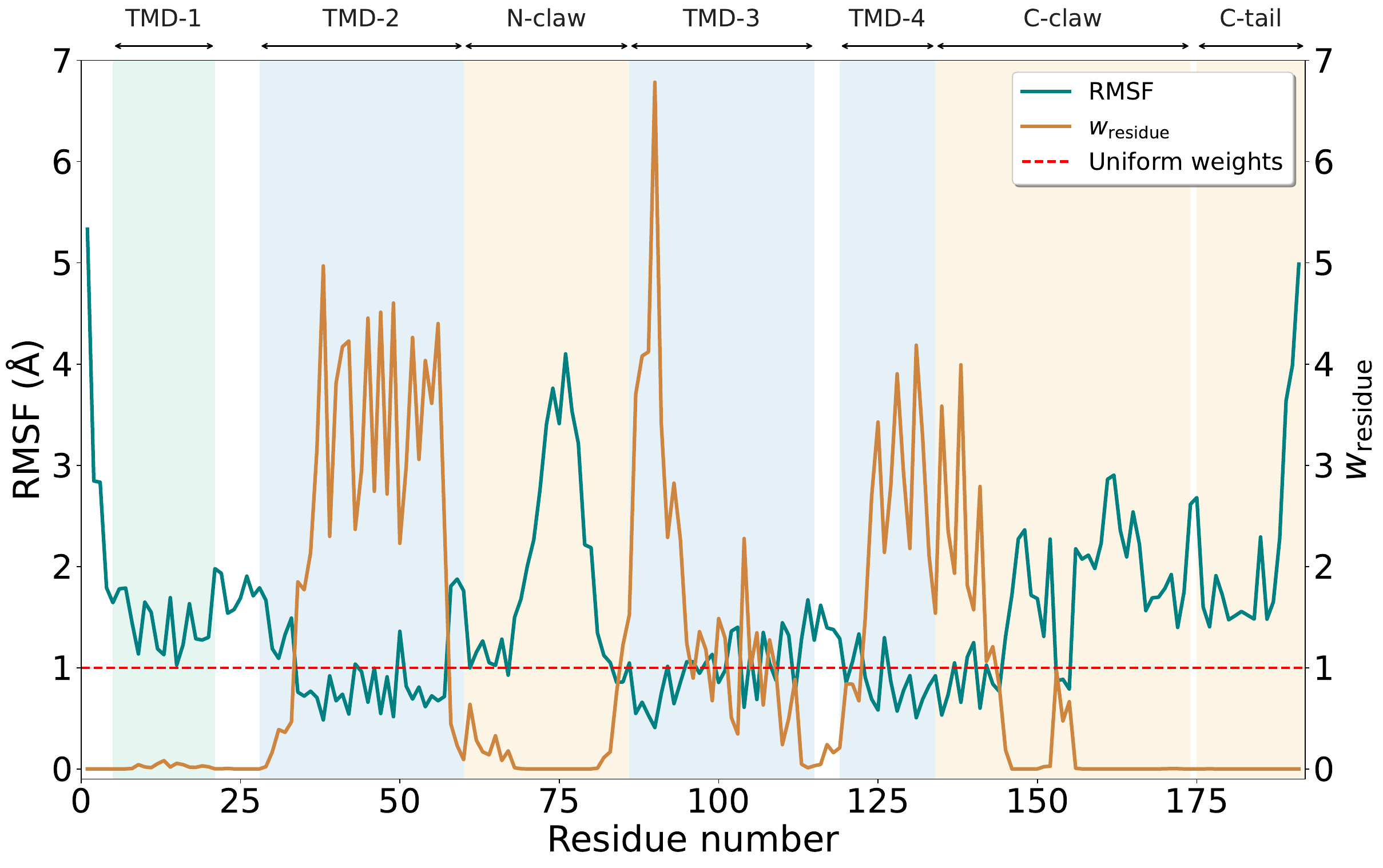}
  \caption{Per-residue RMSF (teal, left axis) and per-residue weight $w_{\rm residue}$ (brown, right axis) at $\sigop = 0.43$~\AA{} for SND3 are plotted as functions of residue number. Domain shading indicates TMD-1 (green), the TM groove TMD-2/TMD-3/TMD-4 (blue), and the mobile N-claw/C-claw/C-tail (orange). Weights and RMSF are anticorrelated, so high-RMSF residues receive exponentially suppressed weights.}
  \label{figS:rmsf_weight}
\end{figure}

\begin{figure}[ht]
  \centering
  \includegraphics[width=\linewidth]{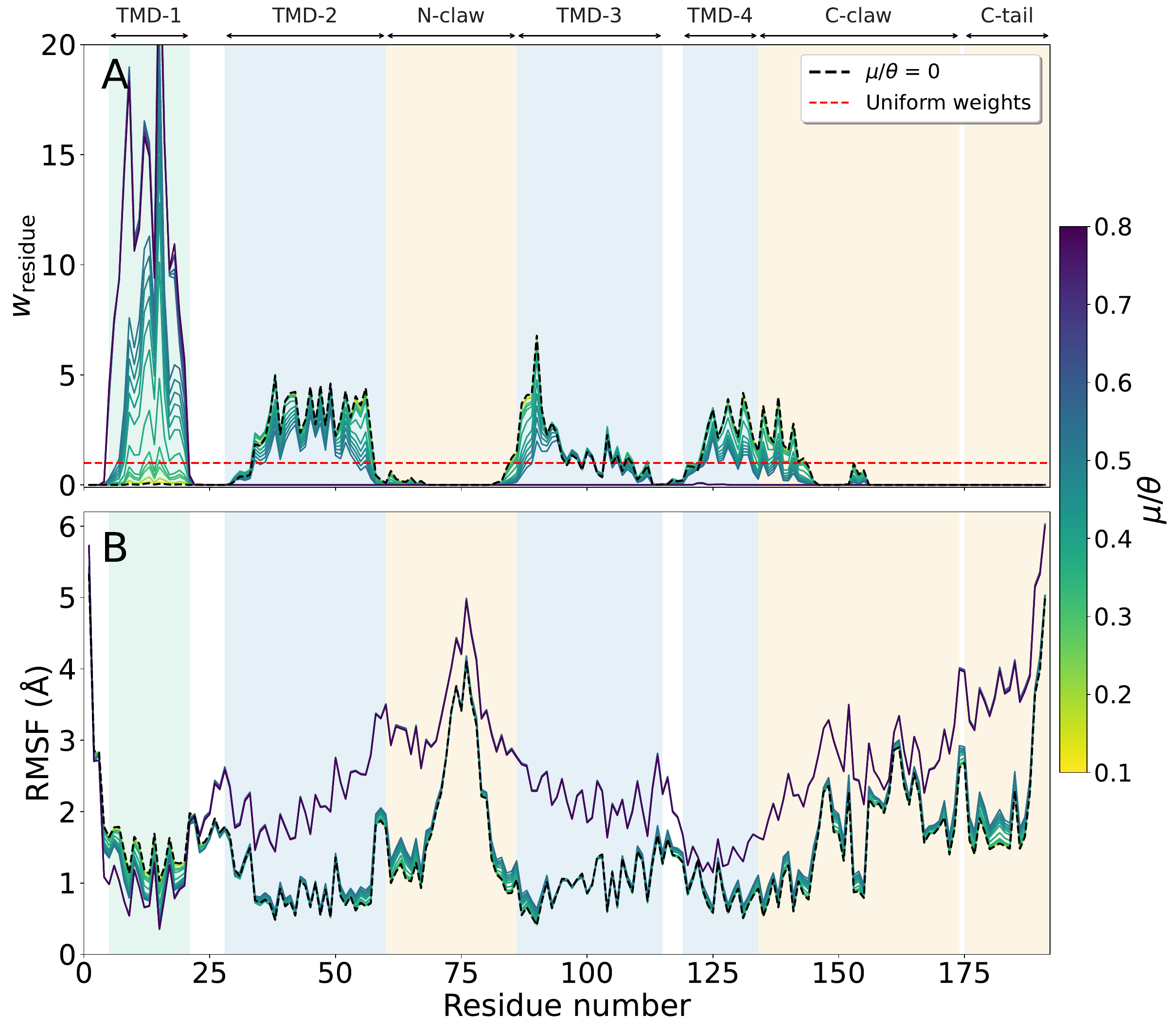}
  \caption{Weight and RMSF profiles for SND3 focused alignment onto TMD-1 are shown for $\mu/\theta = 0$--0.8. (A) Per-residue weights $w_{\rm residue}$ as a function of residue number. The TMD-1 (green shading) weights grow progressively as $\mu/\theta$ increases from the global-alignment baseline (blue) to the operating point (orange, $(\mu/\theta)_\text{op} = 0.53$). (B) Per-residue RMSF as a function of residue number for all $\mu/\theta$ values. The RMSF is insensitive to $\mu/\theta$, confirming that the weight redistribution reflects the focusing constraint rather than changes in atomic fluctuations.}
  \label{figS:focus_profiles}
\end{figure}

\begin{figure}[ht]
  \centering
  \includegraphics[width=\linewidth]{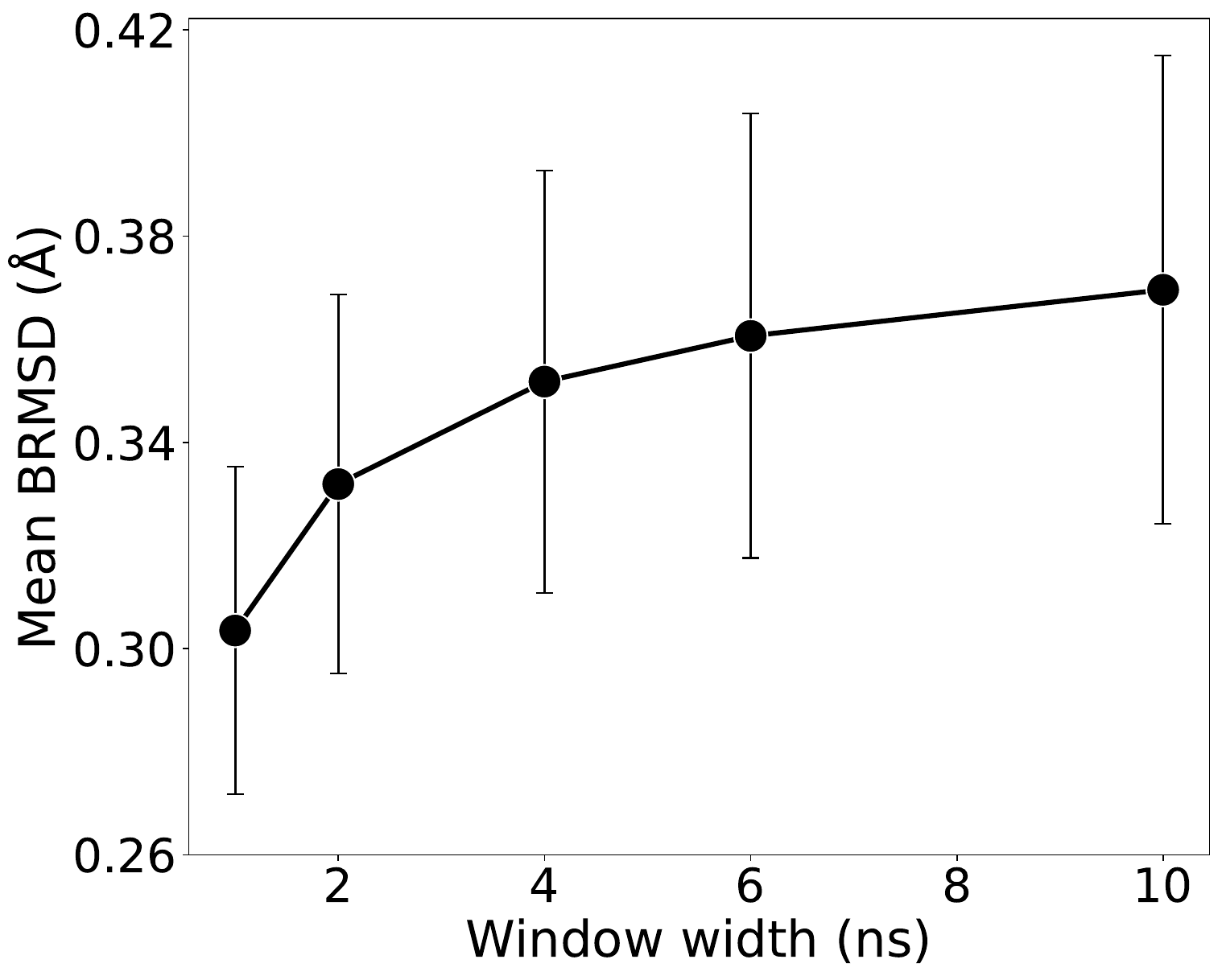}
  \caption{For SND3 smoothing, the mean BRMSD deviation of each raw frame from its locally windowed average is plotted versus total window width ($2 \times$ half-width, ns). Points at full window widths 1, 2, 4, 6, and 10~ns correspond to half-widths $W$ = 5, 10, 20, 30, and 50 frames. The mean deviation increases monotonically from 0.30~\AA{} to 0.37~\AA{} as wider windows produce local averages further from any individual raw frame. The operating point $W_\text{op} = 20$ (4~ns) is marked by a dashed line. Error bars represent standard deviation.}
  \label{figS:smooth_sweep}
\end{figure}

\begin{figure}[ht]
  \centering
  \includegraphics[width=\linewidth]{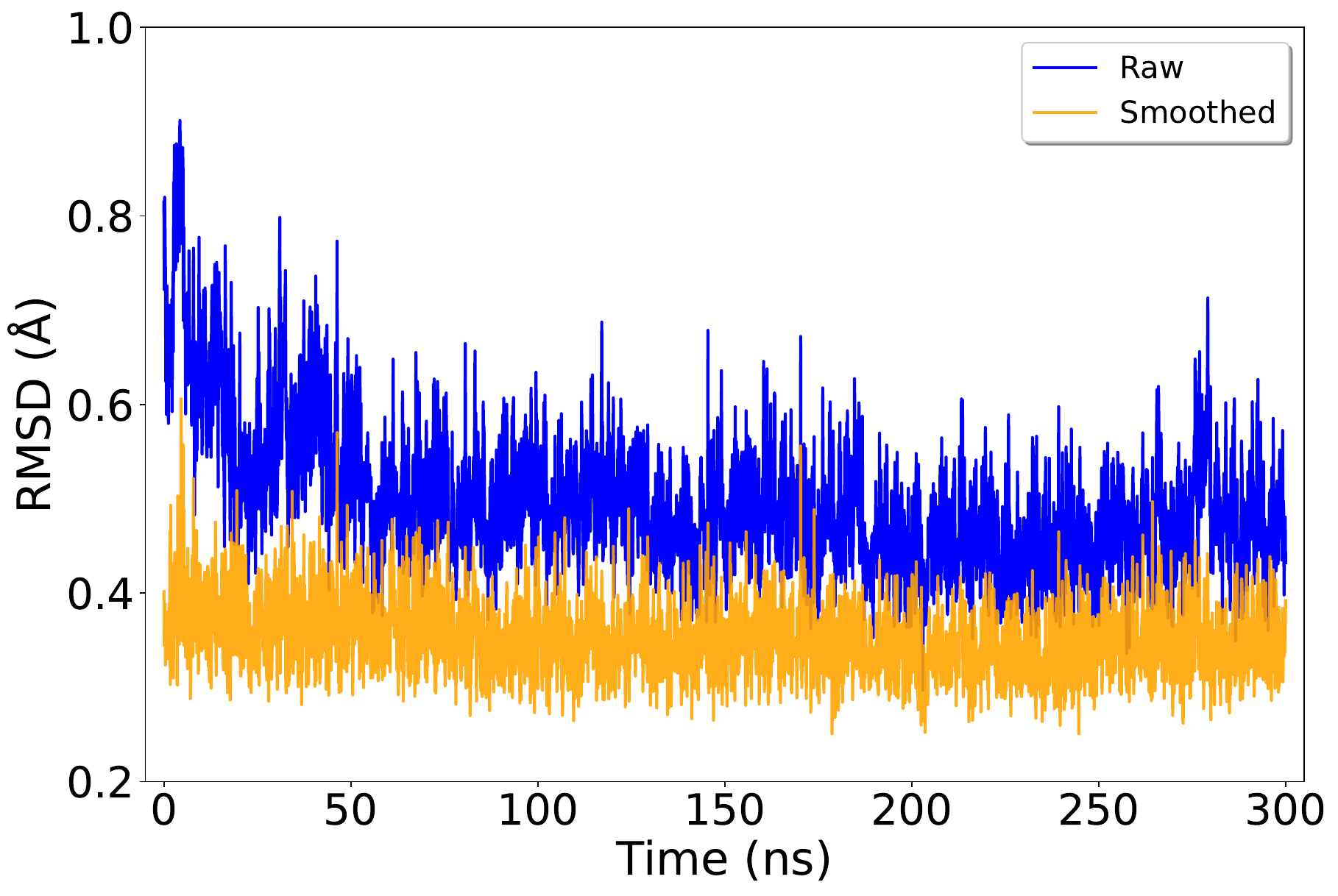}
  \caption{Time series of raw (blue) and smoothed (orange, $W_\text{op} = 20$ frames, 4~ns full triangular window) BRMSD for SND3 over 300~ns. Smoothing reduces the mean from $0.50$ to $0.35$~\AA{} (30\% reduction) and the standard deviation from 0.08 to 0.04~\AA{} (50\% reduction).}
  \label{figS:smooth_ts}
\end{figure}

\begin{figure}[ht]
  \centering
  \includegraphics[width=\linewidth]{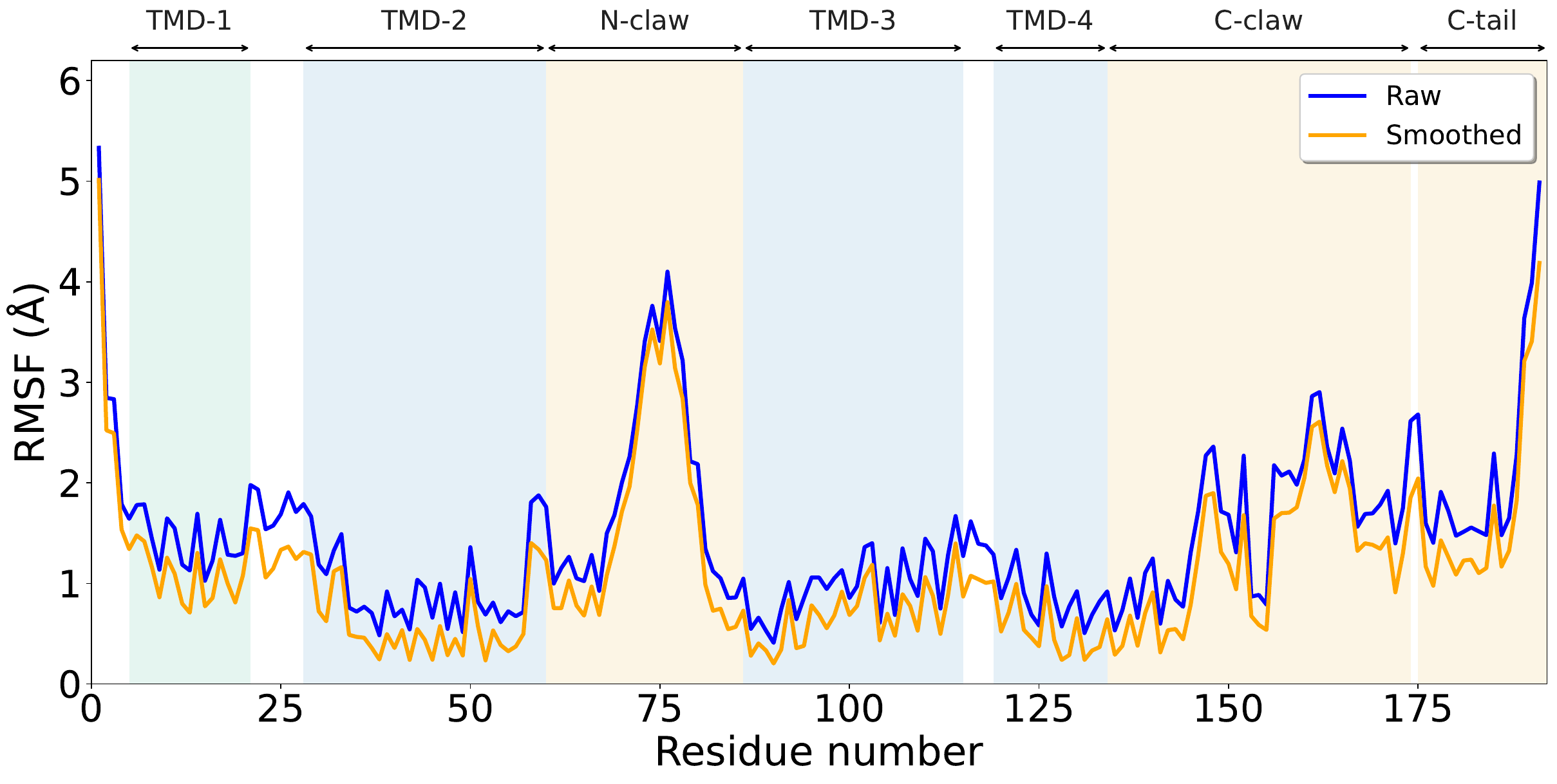}
  \caption{Per-residue RMSF for SND3 before (blue) and after (orange) smoothing at $W_\text{op} = 20$ frames. Smoothing reduces RMSF throughout the chain, and the largest absolute reductions occur in the mobile N-claw, C-claw, and C-tail regions. Domain shading follows the scheme in Fig.~\ref{figS:rmsf_weight}.}
  \label{figS:smooth_rmsf}
\end{figure}

  \begin{figure}[ht]
    \centering
    \includegraphics[width=0.85\linewidth]{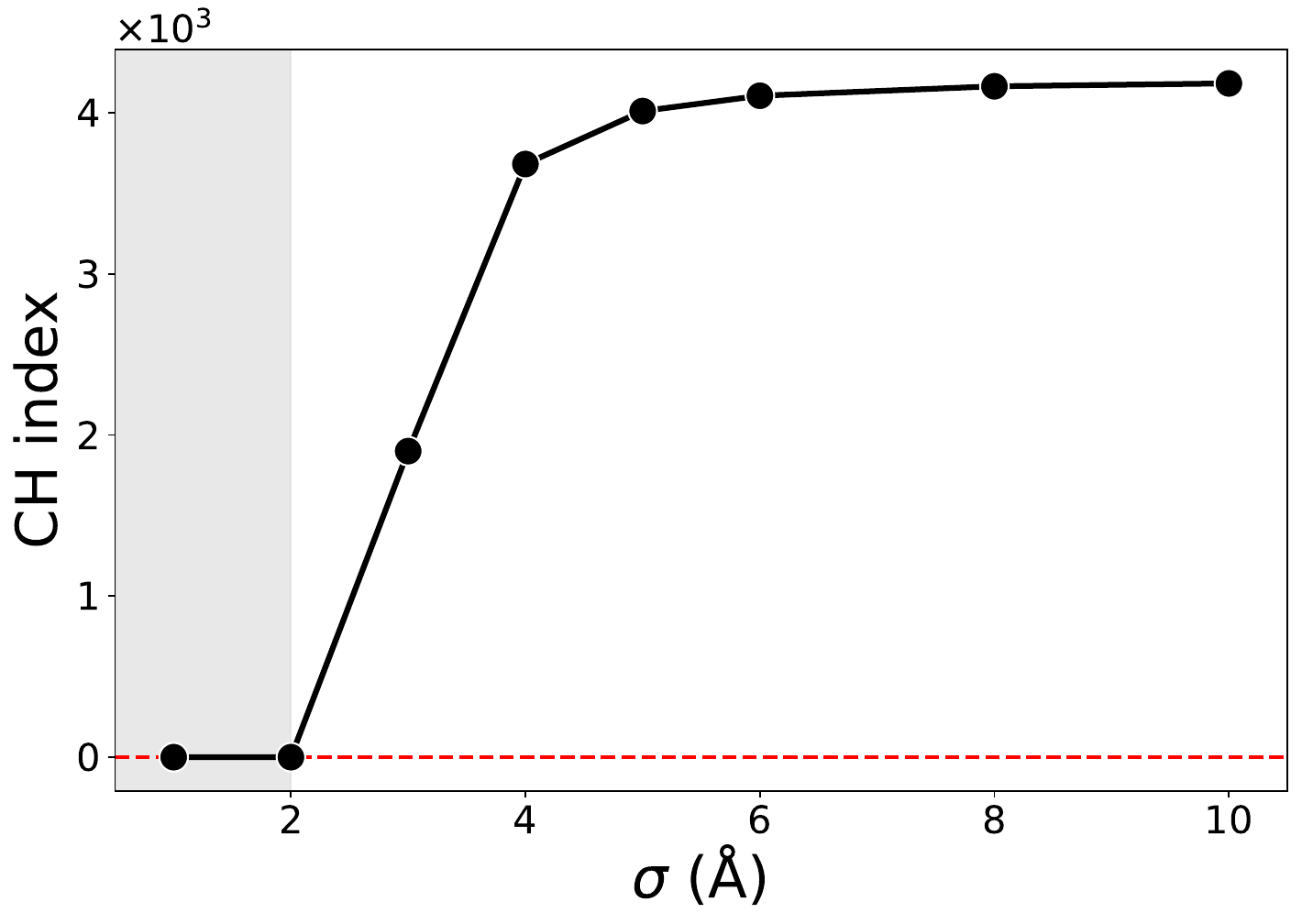}
    \caption{Clustering quality versus the alignment $\sigma$ for the ADK DIMS ensemble ($K = 2$, $\tau = 5.0$~\AA$^2$). The Calinski--Harabasz (CH) index collapses to zero for $\sigma \lesssim 2$~\AA{} (shaded), where weight concentrates on the rigid CORE and the two cluster centres merge. CH index rises sharply at $\sigma=3$–$4$~\AA{} and plateaus for larger $\sigma$.}
    \label{figS:cluster_sigma}
  \end{figure}

\begin{figure}[ht]
  \centering
  \includegraphics[width=\linewidth]{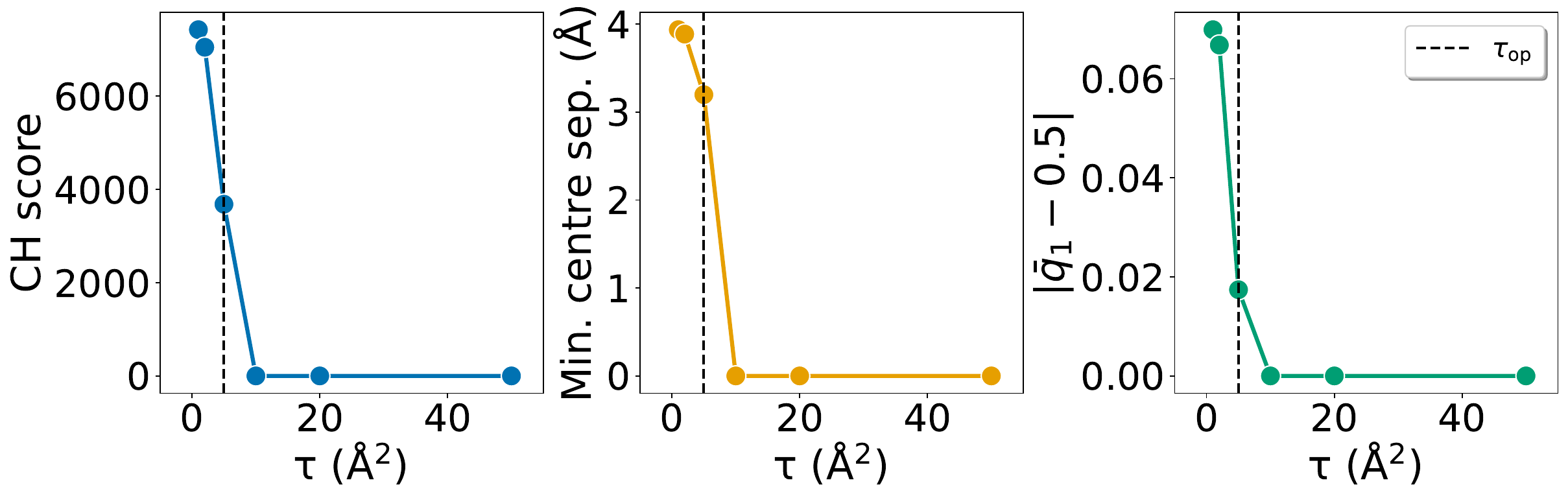}
  \caption{Results of the $\tau$ sweep at $K = 2$ for ADK DIMS clustering. Three panels show the Calinski--Harabasz (CH) index (left), the minimum cluster-centre separation in BRMSD units (centre), and the population imbalance $|\bar{q}_1 - 0.5|$ (right) versus $\tau$. CH and centre separation stay high for $\tau \le 5$~\AA$^2$ and collapse to zero at $\tau \ge 10$~\AA$^2$, where the responsibilities become uniform and the centres merge. The dashed line marks $\taup = 5.0$~\AA$^2$, the largest $\tau$ that preserves discrimination and gives the most balanced partition.}
  \label{figS:tau_sweep}
\end{figure}

\begin{figure}[ht]
  \centering
  \includegraphics[width=\linewidth]{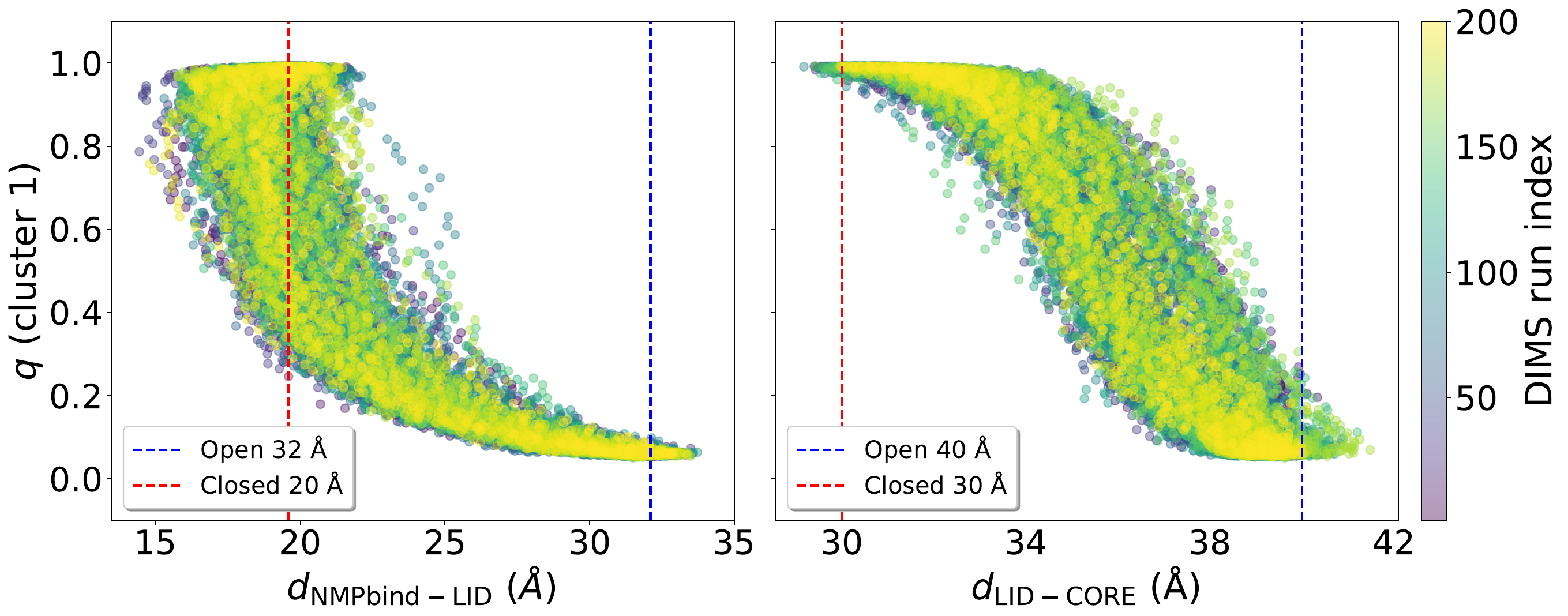}
  \caption{ADK cluster responsibilities $q(\text{cluster 1}|i)$ are plotted versus NMP--LID distance (left) and LID--CORE distance (right), colored by run index. Both order parameters correlate strongly with cluster membership (Pearson $r$ = -0.88 and -0.94, respectively). Vertical lines mark the open (4AKE) and closed (1AKE) crystal-structure reference distances.}
  \label{figS:responsibilities}
\end{figure}

\begin{figure}[ht]
  \centering
  \includegraphics[width=\linewidth]{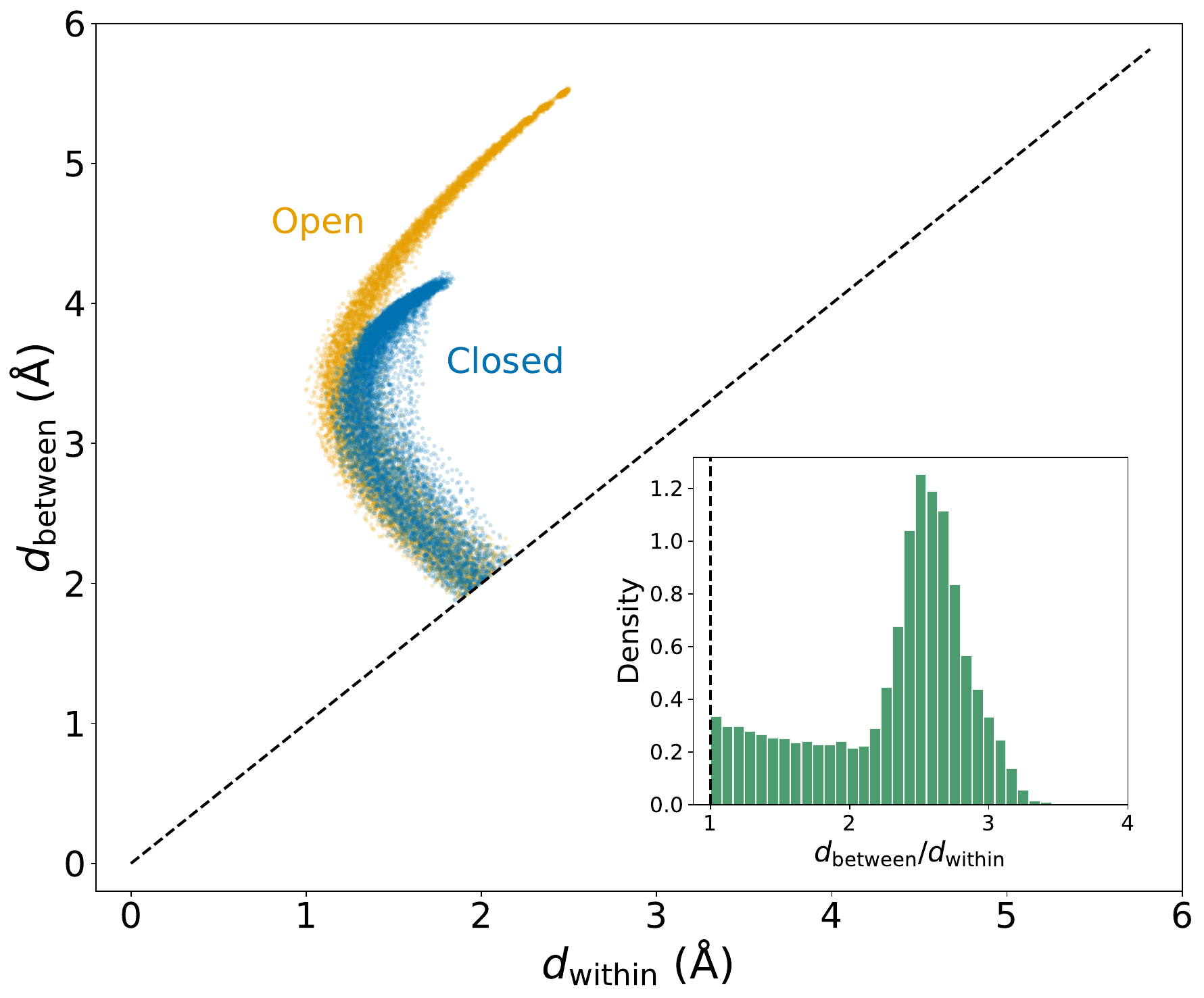}
  \caption{For ADK, within- and between-cluster BRMSD distances are compared in two panels. (Left) Scatter of $d_\text{between}$ versus $d_\text{within}$ colored by cluster assignment (orange: cluster~1, blue: cluster~2), where the dashed diagonal marks $d_\text{between} = d_\text{within}$. All frames satisfy $d_\text{between} > d_\text{within}$, confirming genuine cluster separation. Inset shows the probability density of the ratio $d_\text{between}/d_\text{within}$, with a median ratio of 2.48.}
  \label{figS:separation}
\end{figure}

\begin{figure}[ht]
  \centering
  \includegraphics[width=\linewidth]{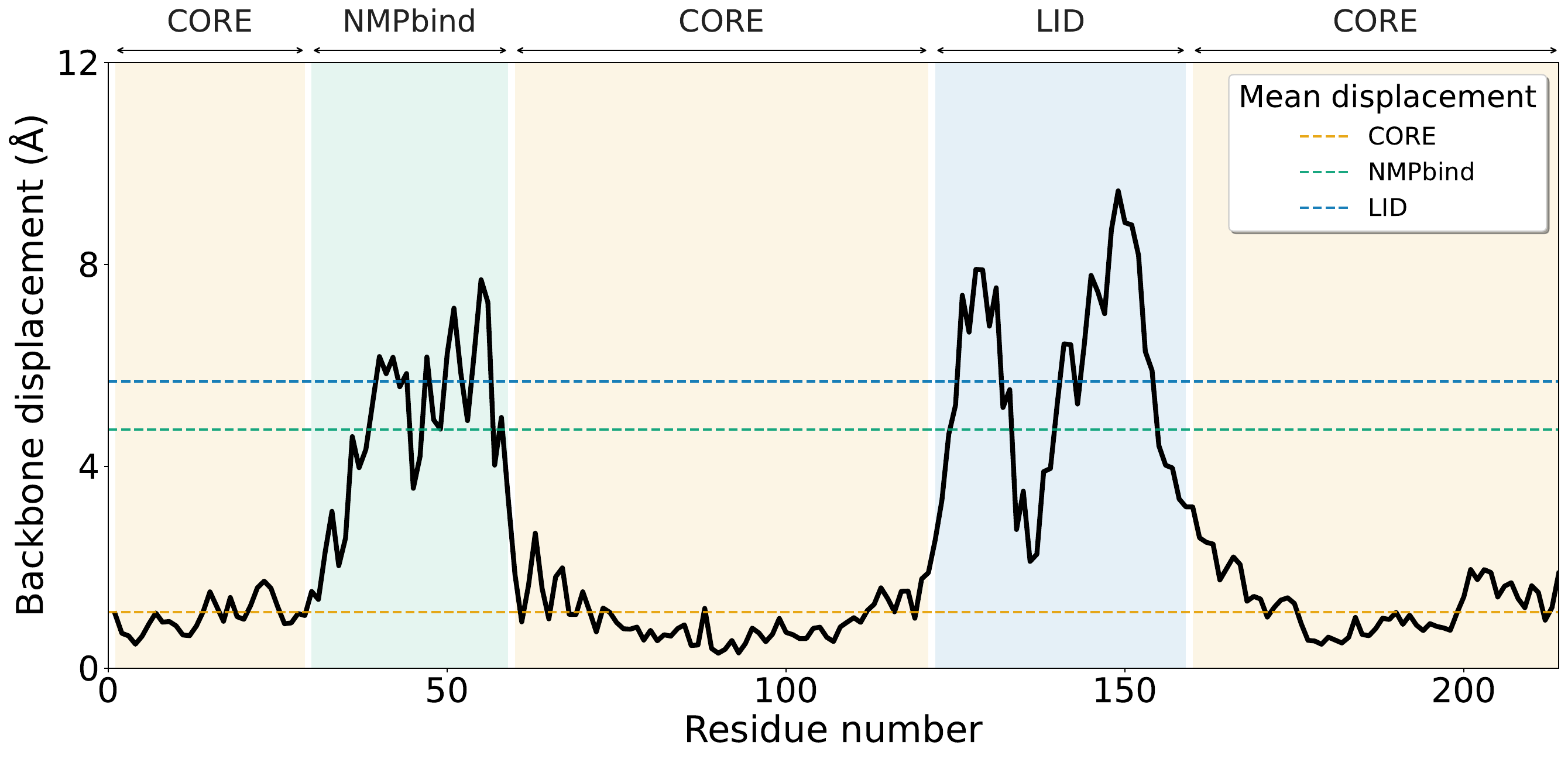}
  \caption{Per-residue displacement between the $K = 2$ cluster centres in ADK is shown as a function of residue number. Mean displacements of the CORE, NMPbind, and LID domains are 1.11~\AA{}, 4.73~\AA{}, and 5.68~\AA{}, respectively. The hierarchy CORE $<$ NMPbind $<$ LID is consistent with the experimentally observed order of domain rigidity.}
  \label{figS:cluster_displacement}
\end{figure}

\begin{figure}[ht]
  \centering
  \includegraphics[width=\linewidth]{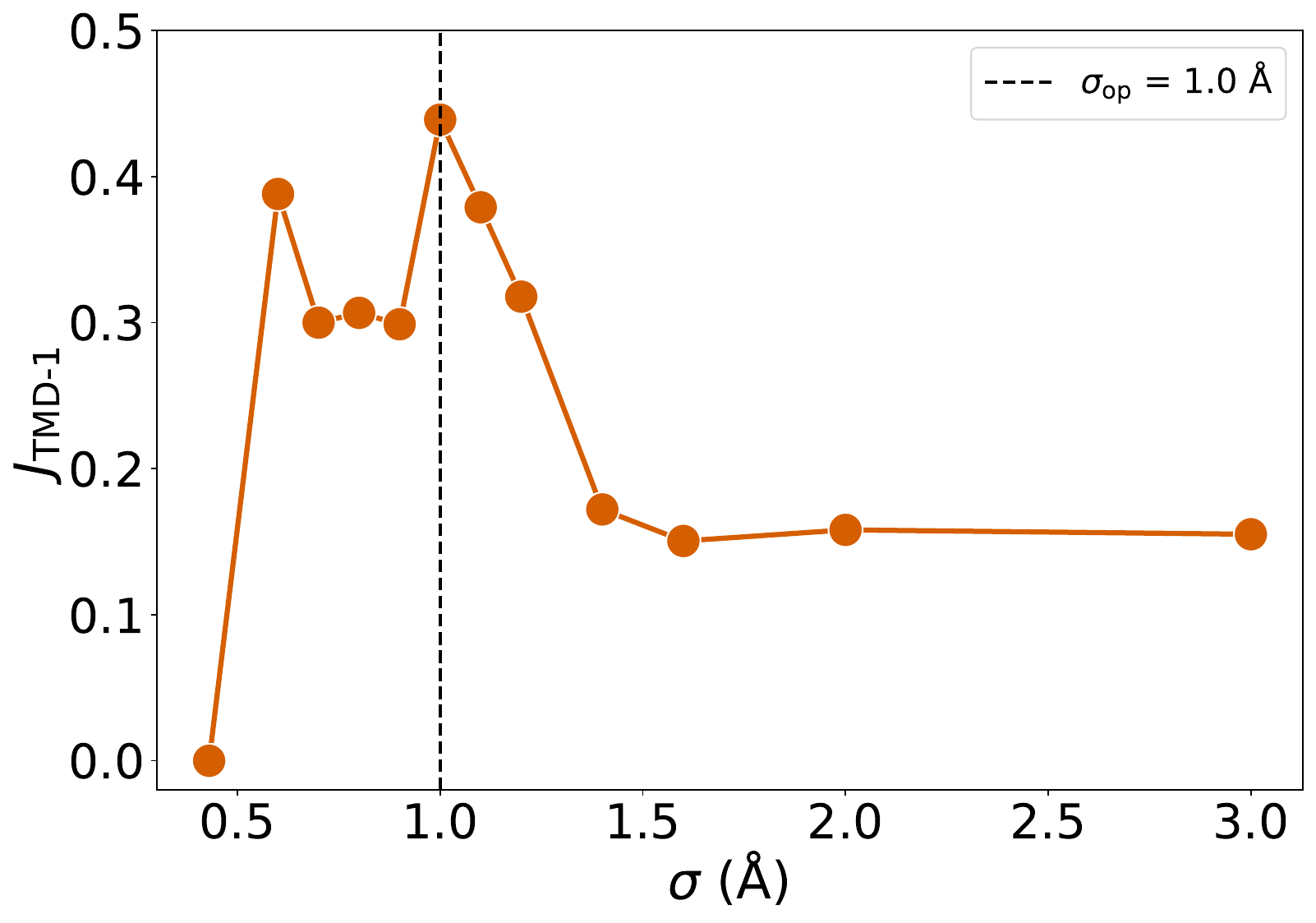}
  \caption{Jaccard index between the best-matching peeling round and the TMD-1 reference, $J_\text{TMD-1}$, as a function of the alignment $\sigma$ for SND3 sequential peeling. The overlap peaks at $\sigma = 1.0$~\AA{} (dashed line) and falls to a low plateau for $\sigma \geq 1.6$~\AA{}, where TMD-1 merges with the groove.}
  \label{figS:snd3_peeling_selection}
\end{figure}

\begin{figure}[ht]
  \centering
  \includegraphics[width=\linewidth]{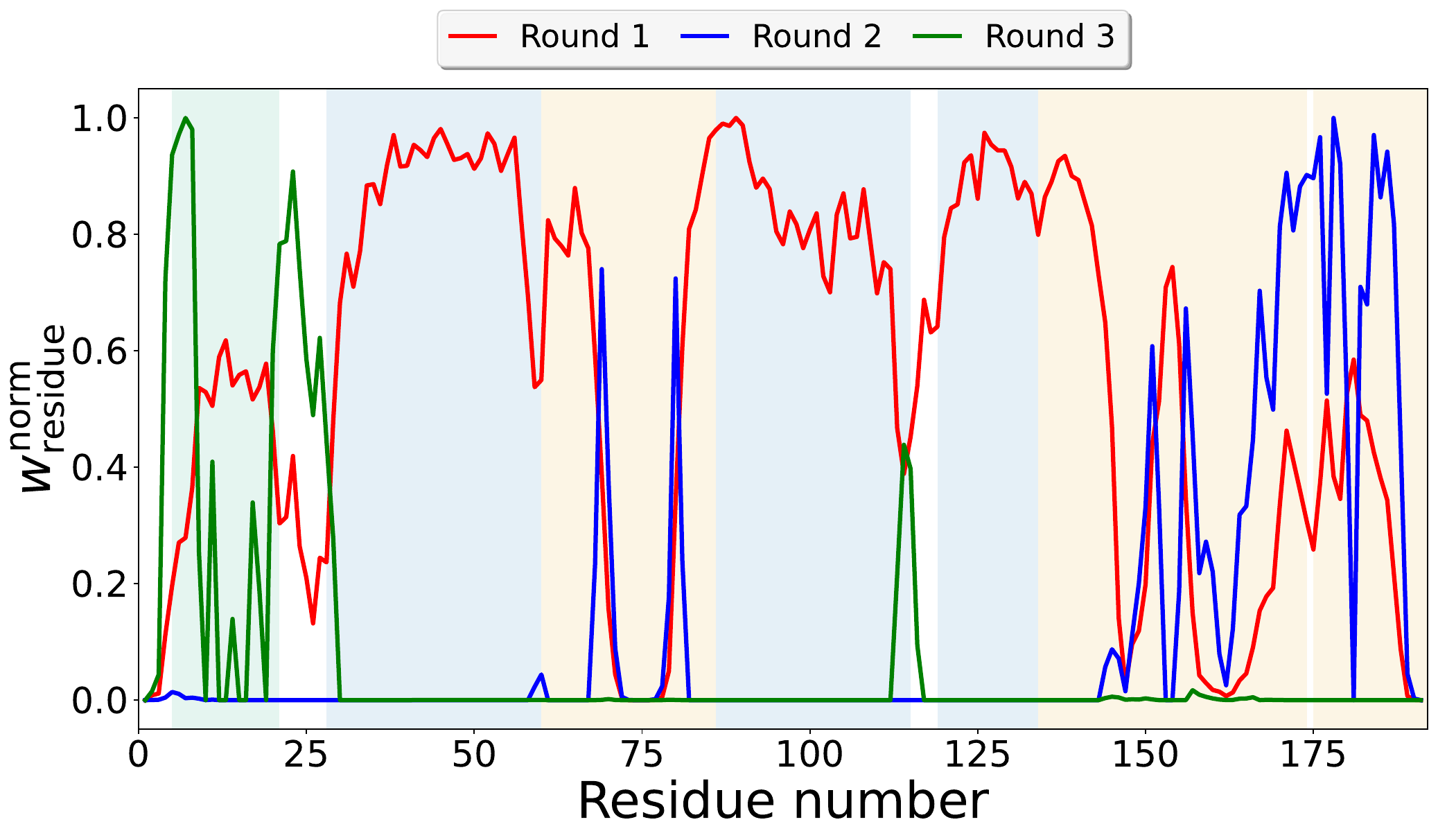}
  \caption{Per-residue weights $w^{\rm norm}_{\rm residue}$ (normalized within each round to its maximum) from SND3 sequential peeling at $\sigma = 1.0$~\AA{}. Round~1 (red) concentrates on the TM groove, Round~2 (blue) on the cytosolic claws, and Round~3 (green) on TMD-1. Domain shading follows Fig.~\ref{figS:rmsf_weight}.}
  \label{figS:snd3_peeling_weights}
\end{figure}

\begin{figure}[ht]
  \centering
  \includegraphics[width=\linewidth]{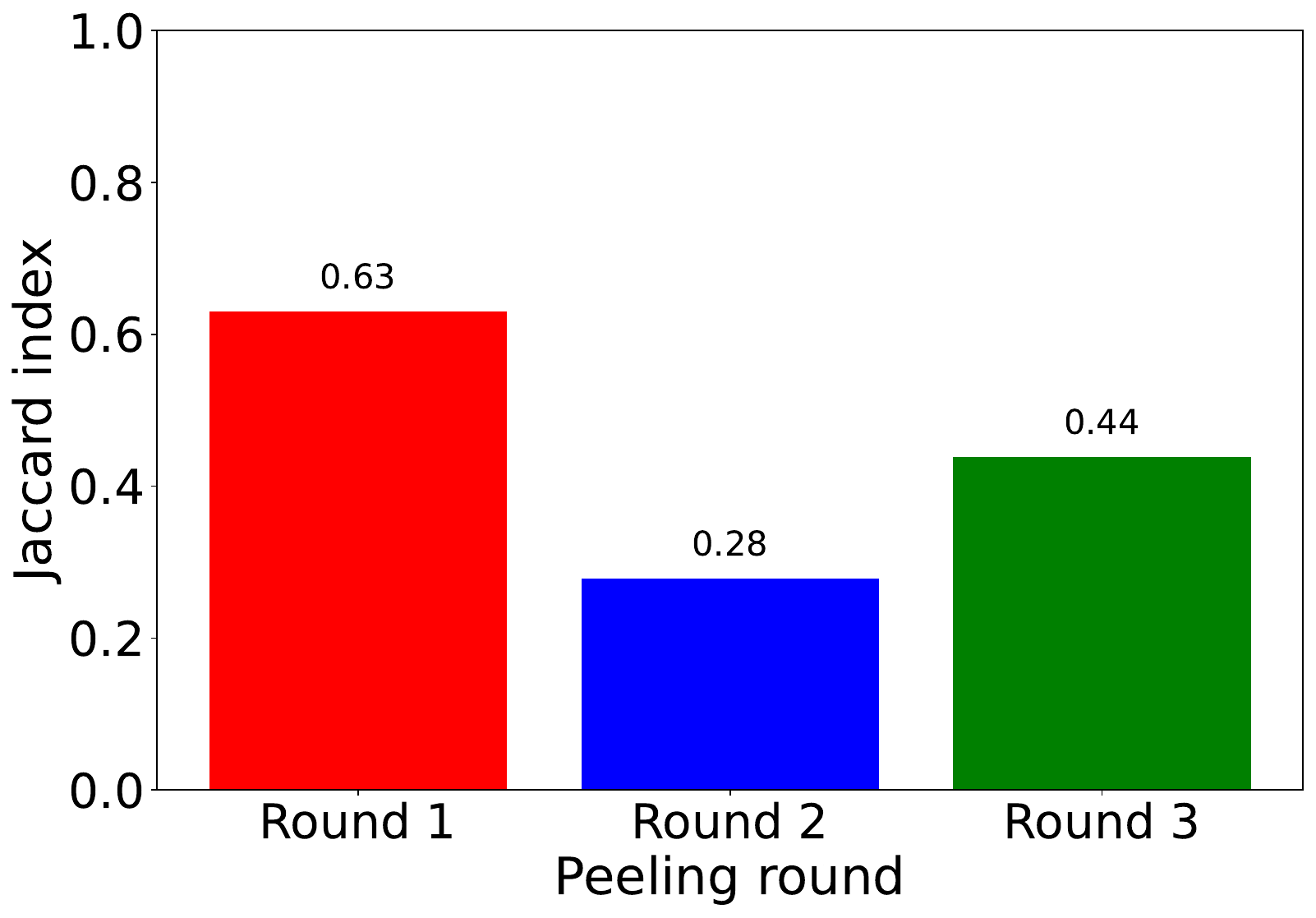}
  \caption{Jaccard index of each SND3 peeling round with its best-matching reference domain at $\sigma = 1.0$~\AA{}. Round~1 recovers the TM groove and Round~3 recovers TMD-1, both distinct from the cytosolic Round~2.}
  \label{figS:snd3_peeling_jaccard}
\end{figure}

\begin{figure}[ht]
  \centering
  \includegraphics[width=\linewidth]{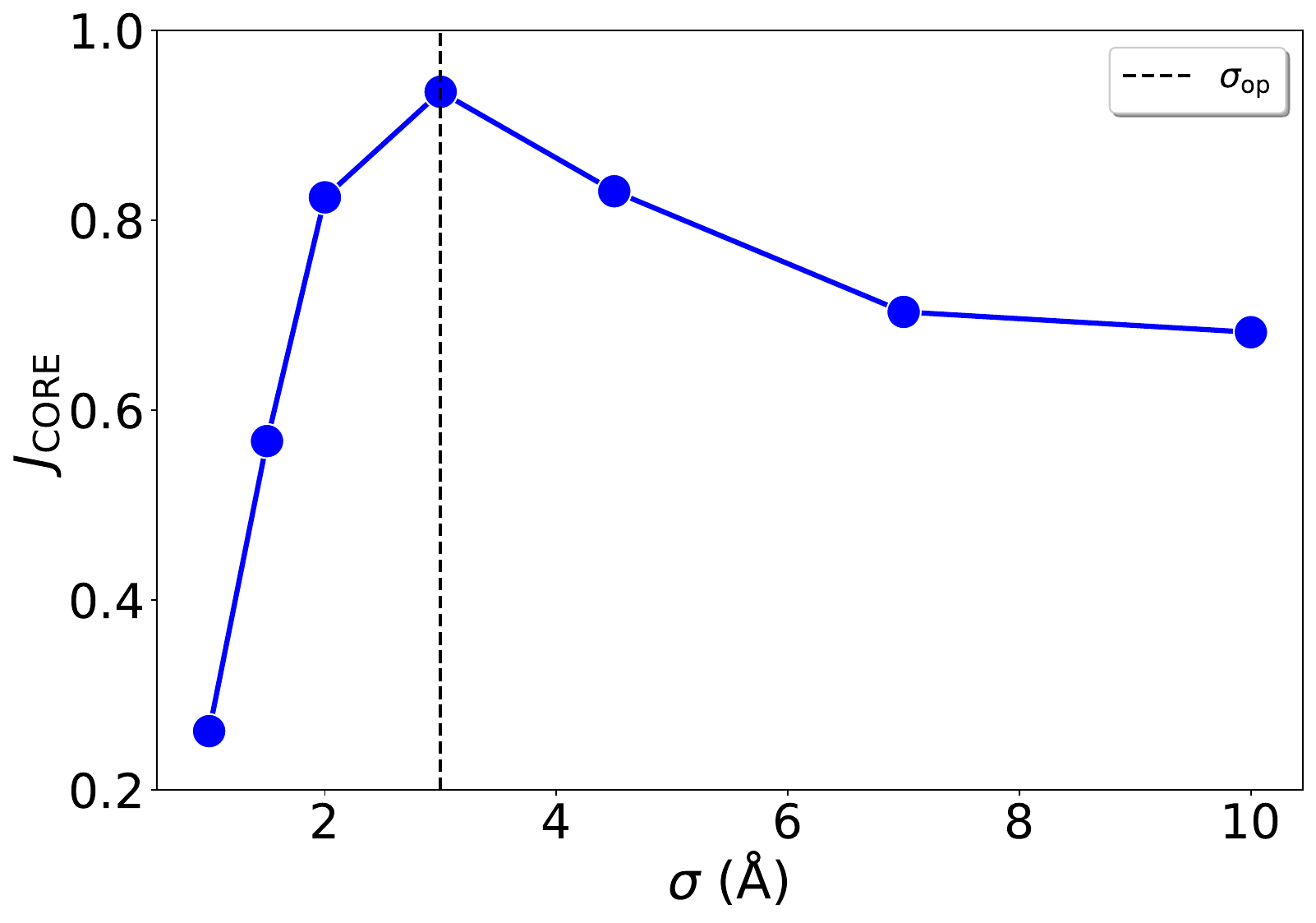}
  \caption{The Jaccard index of the recovered CORE domain, $J_\text{CORE}$, is plotted as a function of $\sigma$ for the $K = 1$ sweep on the full 856-atom ADK FRODA pool. $J_\text{CORE}$ peaks at $\sigop = 3.0$~\AA{} (dashed line, $J_\text{CORE} = 0.94$) and decays at both larger and smaller $\sigma$, defining the operating point used for sequential peeling.}
  \label{figS:adk_sigma_sweep}
\end{figure}

\begin{figure}[ht]
  \centering
  \includegraphics[width=\linewidth]{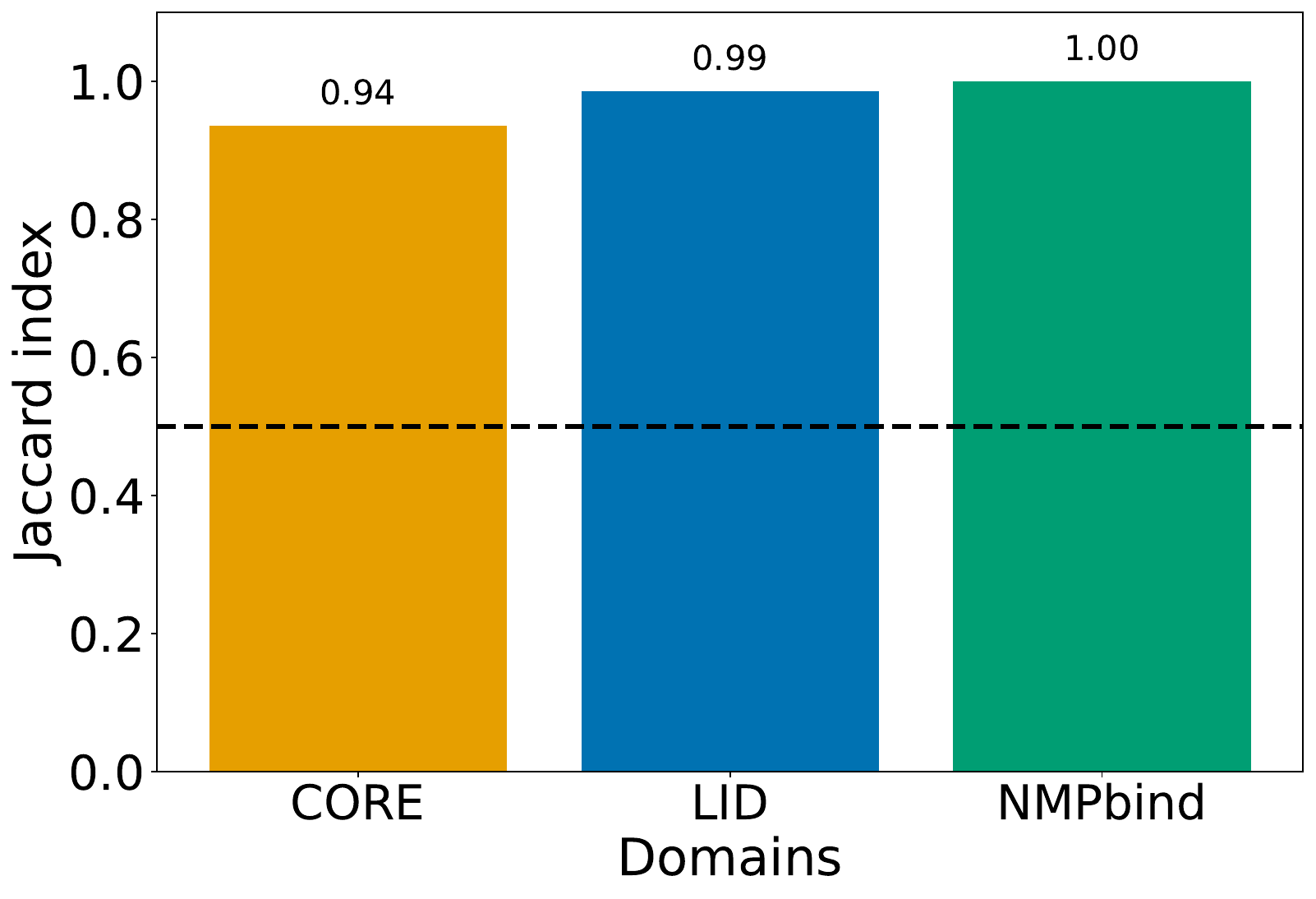}
  \caption{The Jaccard index for each of the three sequential-peeling steps of rigid domain detection for ADK is shown, based on 28,282 FRODA backbone frames ($\sigop = 3.0$~\AA{}, $N = 856$). The recovered Jaccard indices are $J_{\rm CORE} = 0.94$ (Step~1, 856 atoms), $J_{\rm LID} = 0.99$ (Step~2, 240 non-CORE atoms), and $J_{\rm NMPbind} = 1.00$ (Step~3, 96 remaining atoms). The dashed line marks the $J = 0.5$ recovery threshold.}
  \label{figS:adk_jaccard}
\end{figure}

\begin{figure}[ht]
  \centering
  \includegraphics[width=\linewidth]{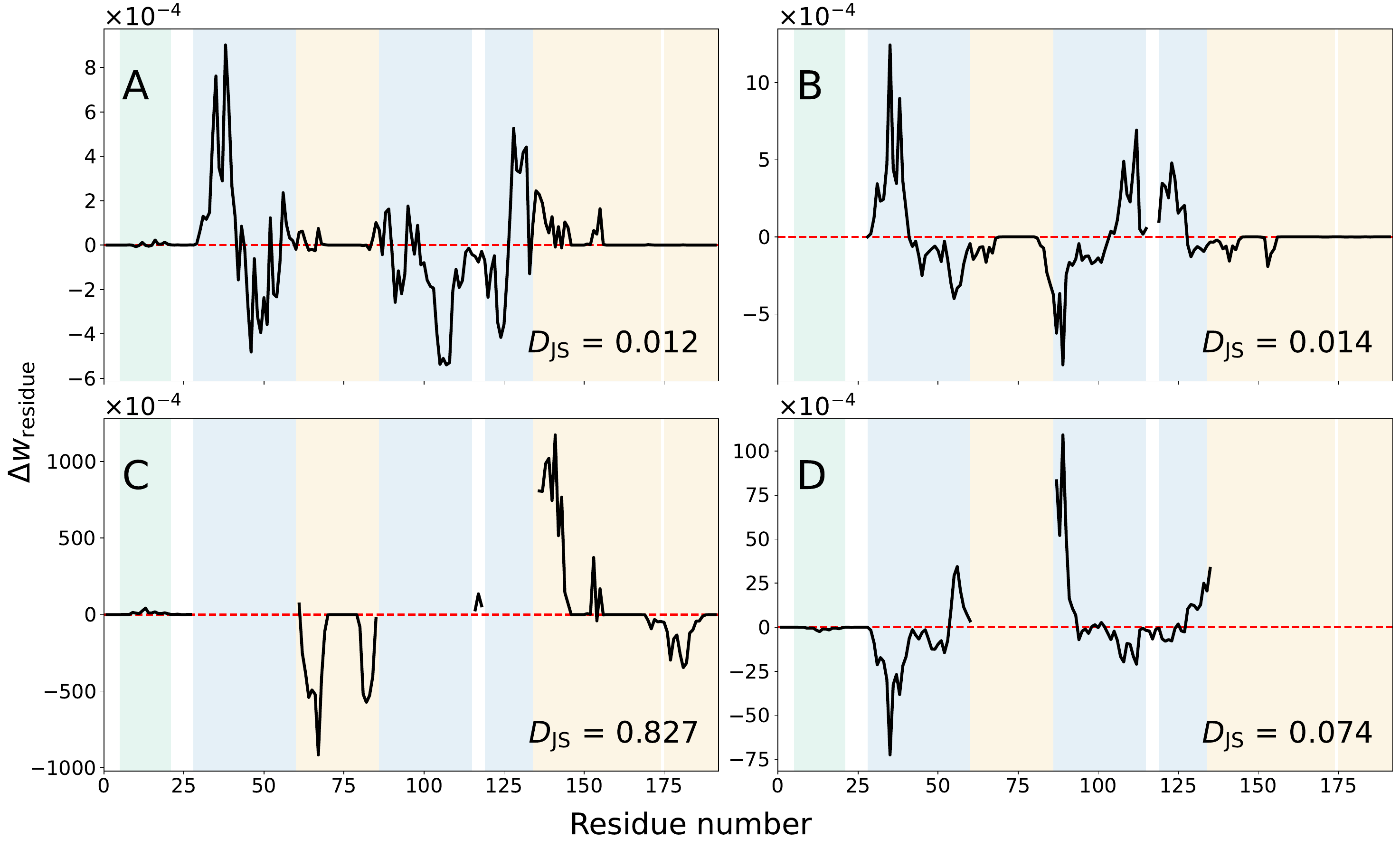}
  \caption{Consistency of the SND3 alignment weights at $\sigop = 0.43$~\AA{}. Each panel shows the per-residue weight difference $\Delta w_{\rm residue}$ between two optimizations. (A) $M$-invariance: all 3001 frames versus every other frame (1501 frames), $D_\text{JS} = 0.012$. (B)--(D) Pool-invariance: a whole region is removed from the selection and the weights are re-optimized on the retained atoms, compared to the corresponding atoms of the full solution. Removing TMD-1 (B, $D_\text{JS} = 0.014$) or the cytosolic claws (D, $D_\text{JS} = 0.074$) leaves the retained weights unchanged, whereas removing the rigid TM groove (C, $D_\text{JS} = 0.827$) forces a different solution. Domain shading follows Fig.~\ref{figS:rmsf_weight}.}
  \label{figS:consistency}
\end{figure}

\begin{figure}[ht]
  \centering
  \includegraphics[width=\linewidth]{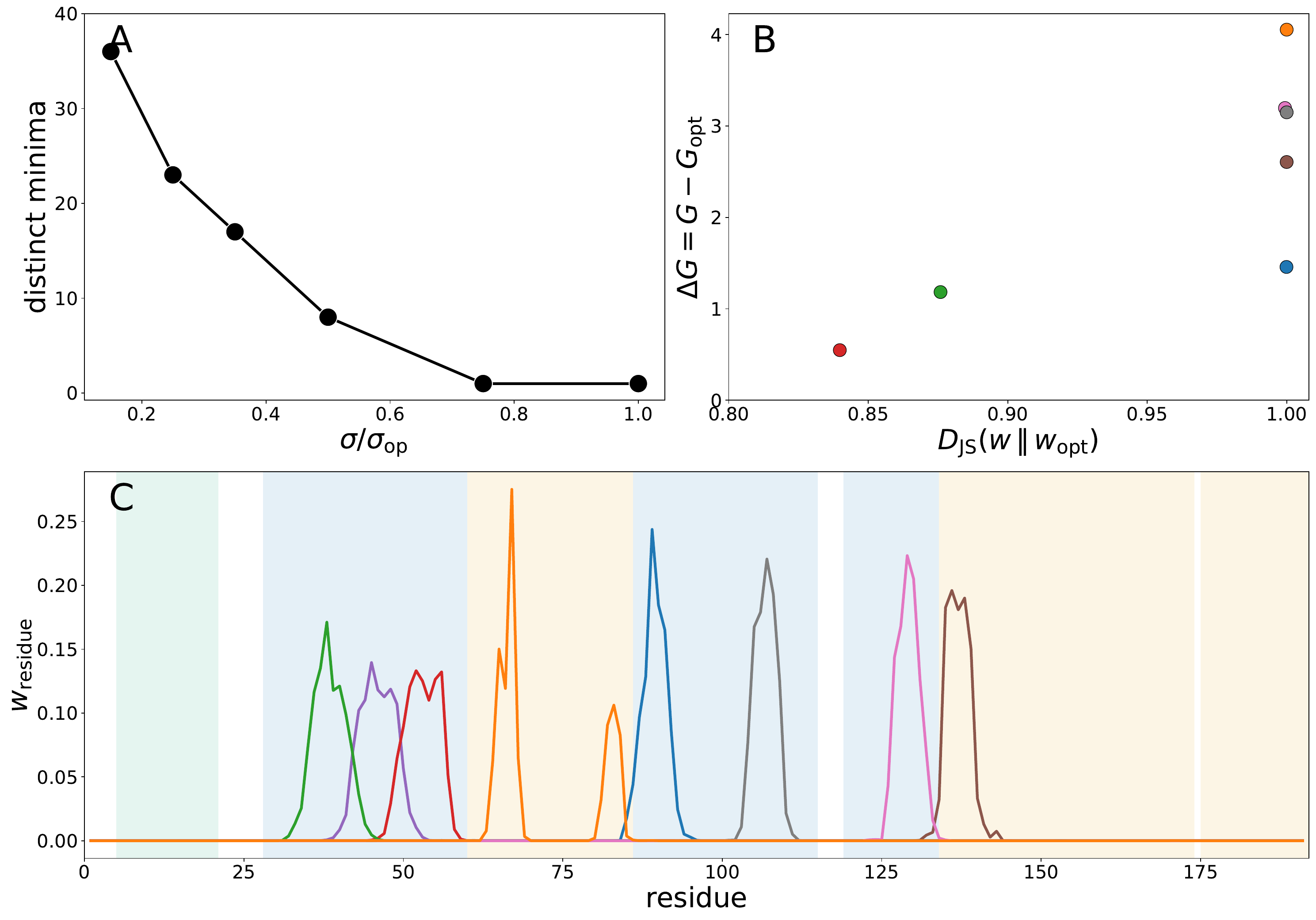}
  \caption{Fixed-point weight solutions of the SND3 alignment ($\theta = M\sigma^2$), obtained from 40 localized (Gaussian-bump) initializations clustered by Jensen--Shannon distance. (A) Number of distinct minima as a function of $\sigma/\sigop$. At $\sigop$, there is a single solution. It fragments into many as $\sigma$ decreases. (B) Free-energy offset $\Delta G = G - G_\text{opt}$ versus $D_\text{JS}(\boldsymbol{w}\,\|\,\boldsymbol{w}_\text{opt})$ at $0.5\,\sigop$, showing the 7 higher-$G$ alternative minima. The global solution ($\Delta G=0$) is not shown. All alternatives have higher $G$ than the global minimum. (C) Per-residue weight profiles of the 8 minima at $0.5\,\sigop$, each localized on a different rigid sub-structure.}
  \label{figS:minima}
\end{figure}

\clearpage
\subsection*{Supplementary Movies}

\noindent\textbf{Movie~S1.} Raw (left) and BRMSD-aligned (right) SND3 trajectories are shown side by side ($\sigop = 0.43$~\AA{}). The lighter colors represent the weight-concentrated transmembrane groove.

\noindent\textbf{Movie~S2.} Raw (left) and smoothed (right) SND3 trajectories are shown side by side ($W_\text{op} = 20$ frames, 4~ns full triangular window, $\sigop = 0.43$~\AA{}). Sub-nanosecond fluctuations are suppressed while slow movements are preserved.

\begin{tocentry}
\begin{center}
    \includegraphics[width=\linewidth]{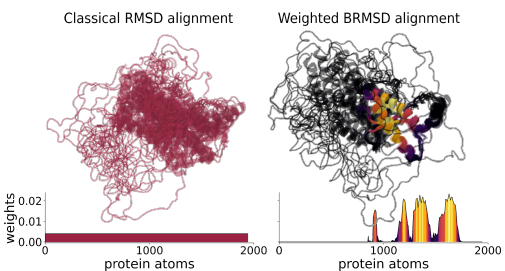}
\end{center}
\end{tocentry}

\bibliography{ref}

\end{document}